# Atmospheric dynamics of IR-active particles released from Mars' surface.

Mark I. Richardson[1,*], Samaneh Ansari[2], Bowen Fan[3,4], Ramses Ramirez[5], Hooman Mohseni[2], Michael A. Mischna[6], Michael H. Hecht[7], Liam J. Steele[3,8], Felix Sharipov[9], Edwin S. Kite[3,*]

1. Aeolis Research, Chandler, AZ.
2. Northwestern University, Evanston, IL.
3. University of Chicago, Chicago, IL.
4. Now at: Yale University, New Haven, CT.
5. University of Central Florida, Orlando, FL.
6. Jet Propulsion Laboratory, California Institute of Technology, Pasadena, CA.
7. MIT Haystack Observatory, Westford, MA.
8. Now at: European Center for Medium-Range Weather Forecasts, Reading, UK.
9. Universidade Federal do Paraná, Curitiba, Paraná, Brazil.

*Corresponding authors, mir@aeolisresearch.com and kite@uchicago.edu

**Key points:**

- Using a plume-tracking climate model, we investigate radiative-dynamical feedback from release of two particle compositions at Mars' surface.
- For a single, constant source, global particle abundance saturates in no more than 4 Mars Years, enabling Mars surface and atmospheric warming.
- Self-lofting helps particles rise and spread, and an enhanced Hadley cell aids global mixing.

**Plain language summary.**

Both aerosols and gases have been proposed as ways of warming Mars. For the aerosol method, particles released locally must disperse globally. We address this question for the first time using a plume-tracking, dry Martian global atmospheric model. We find that for a single, constant source, global particle abundance saturates in <4 Mars Years, enabling Mars surface and atmospheric warming. Self-lofting helps particles rise and spread, and stronger winds help global mixing. Warming is strong, but varies depending on model set-up (stronger in 3-D models than in 1-D models). Many open atmospheric science questions remain, including the role of agglomeration, dry-deposition rate uncertainty, and modeling water cycle feedbacks.

**Abstract.**

Surface release of radiatively active particles, with high infrared- (IR-)to-visible extinction ratios, has been proposed as a method of warming Mars. However, to warm Mars using aerosols, particles released locally must disperse globally. Here we provide an initial reference study in a plume tracking, dry Martian atmospheric model to address this question. The winds that transport aerosols respond to the aerosol's IR forcing, implying strong radiative-dynamical feedbacks (RDF). We investigate RDF from surface release of two particle compositions: carbon (graphene) and metal (Al). Self-lofting helps particles rise and spread locally and regionally, and the Hadley cell strengthens under warming, aiding latitudinal mixing. Within our model, Mars RDF enable engineered-aerosol warming. Warming is slightly greater for three-dimensional *vs*. 1D-models and also depends on spectral resolution of radiative transfer. We assess implications for Mars warming. Many open atmospheric science questions remain, including the role of agglomeration, dry-deposition rate uncertainty, and modeling water cycle feedbacks.

## 1. Introduction.

Mars is a cold desert, with a weak greenhouse effect (+5 K from its 6 mbar $CO_2$ atmosphere), abundant frozen $H_2O$ (Carr & Head 2015), and 7-100 mbar of sequestered volatilizable $CO_2$ (Bierson et al. 2016, Buhler & Piqueux 2020). It has been debated for decades whether or not humans should warm Mars for the purposes of human habitation, with arguments both for (McKay 1990) and against (Marshall 1993). If we decide in the future to modify Mars' environment (Sagan 1973), there is consensus that warming Mars by >30 K would be needed for seasonal meltwater (e.g., Averner et al. 1976, McKay et al. 1991, Pollack & Sagan 1993, Graham 2004, Marinova et al. 2005); an up-to-date review is DeBenedictis et al. (2025). While such warming long seemed impractical (Jakosky & Edwards 2018), recent developments suggest new Mars-warming methods (e.g., Wordsworth et al. 2019), including release of IR-active particles (Ansari et al. 2024). However, prior work on warming with IR-active particles assumed static aerosol distributions based on natural dust aerosol distribution (Ansari et al. 2024) rather than modeling the dynamical behavior of engineered aerosol.

Aerosols are an important radiative agent in planetary atmospheres (Yung & DeMore 1999, Kahre et al. 2017). Previous studies have examined aerosol feedbacks on Mars for natural particles, from local "solar escalator" effects (Spiga et al. 2013, Daerden et al. 2015, Batterson et al. 2023) to larger scales (e.g., Murphy et al., 1993, Wilson and Hamilton, 1996, Newman et al., 2002, Rafkin 2009, Kahre et al. 2015, Gebhardt et al. 2021, Urata et al. 2025) and on Earth for smoke, volcanic ash, and other applications (De Laat et al. 2012, Yu et al. 2019, Khaykin et al. 2020, Visioni et al. 2020, Gao et al. 2021, Ohneiser et al. 2023). However, the engineered aerosols considered for warming Mars have (by design) much stronger interaction with thermal-IR emission than with sunlight, whereas previous work has focused on self-lofting driven by absorption of sunlight. Ansari et al. (2024) considered IR-active particles but assumed a static particle distribution. Given that RDF will be strong for IR-active particles released from Mars' surface, key atmospheric-dynamics questions remain: How high are aerosols lofted? How widely do they disperse? How do they couple with atmospheric circulation locally and globally? How long until steady state is reached? Does release location matter? Any Mars-warming method would face problems of manufacturing and environmental impact (*e.g.,* both potential impact on humans and/or introduced plant life, and on, *e.g*., planetary climate via albedo changes) as well, but the radiative-dynamical questions must be addressed first. To this end, we added radiatively active engineered nanoparticle tracking to the Mars Weather Research and Forecasting (MarsWRF) code (Richardson et al. 2007, Toigo et al. 2012), building on previous work tracking methane plumes (Mischna et al. 2011, Luo et al. 2021) (Methods). We specifically focus in this initial study on the simplest case of non-coagulating particles in a dry atmosphere released from a single, continuous source. This is done to minimize process complexity for this initial study and is not due to limitations with the model. More complex modeling will be necessary in the future to more completely address realistic release scenarios, particulate microphysical behavior, and climate physics. Instead, this study provides the necessary next step from idealized and/or 1D work that has been undertaken so far and provides a reference baseline for interpretation of future, more complex simulations.

## 2. Methods.

We carry out separate numerical experiments that each consider one of two particle types for Mars warming: carbon (graphene disks) and metal (Al rods). Both target Mars' thermal infrared (TIR) windows at ~10 μm and ~20 μm via different mechanisms. Few-hundred-nm diameter graphene disks resonantly absorb TIR (Fang et al. 2014). To cover both spectral windows (Fig. S3), in the graphene trials we used 16 parts (by number) 250 nm diameter disks to one part 1000 nm diameter disks. Graphene is translucent to visible light, though it requires potentially costly modification for optimal performance (Supporting Information). Here, we adjust the electronic properties of the simulated graphene to give resonance height and width similar to Al rods. This implicitly assumes graphene is modified ("doped") to increase the carrier concentration (Supporting Information). The Al rods (60 nm-diameter, 8 μm long, one-eighth the weight of the particles considered in Ansari et al. 2024) both absorb and scatter thermal IR (Forget & Pierrehumbert 1997). Either or both particle types might require anti-stick coatings, which are not modeled here. The goal is to examine plausible particles that might be produced: the chosen particles are not optimized for warming. As such, this study yields potentially realistic heating. It does not represent an upper limit on heating.

In our 3D (MarsWRF) modeling, we simulated continuous near-surface release of IR-active particles at both a northern-hemisphere midlatitude site (202°E 40°N, Arcadia Planitia), and separately at an equatorial site (135°E 0°N, Elysium Planitia), expressing release rate in liters/second (L/s) of solid aerosol (ranging from 0-60 L/s, Table S3). Particles can settle at the surface and we do not include any re-lofting mechanism. For natural dust, we imposed a time-varying background of radiatively active natural dust derived from an observational database (Montabone et al. 2015) for Mars Year 30, corresponding to a relatively storm-free scenario. The vertical distribution was further adjusted to that observed by Mars Climate Sounder (Kleinböhl et al., 2024). Dust was not allowed to interact with the engineered particles. Particle agglomeration is not included in the model, which is a significant limitation (Discussion). Gravitational settling speeds are obtained from analytical free molecular calculations for the appropriate shapes (Supporting Information). We also set the dry deposition velocity in the lowermost model layer, which is ~9 m thick, to a minimum of 0.03 cm/s, to represent the effect of diffusive removal of small particles (Supporting Information). For comparison, the empirical exponential decay timescale for natural dust optical depth is ~50 sols (Kahre et al. 2017).

We also used a 1D radiative-convective equilibrium model with Mars-average insolation (e.g., Ramirez 2017) and a Single Column Model (SCM) version of WRF as a cross-check on warming. Results are shown in Figs. S14-S15 and Table S4. The warming in 1D and 3D models differ (as do baseline global temperatures with no nanoparticle opacity), even when the same radiative transfer scheme is used (WRF SCM *vs.* GCM). The differences can be traced to differences in the vertical temperature profiles (and specifically temperature assumptions at the tropopause), the spectral resolution for radiative transfer, and the treatment of the atmosphere as a 3D system (see Supplementary Materials).

**3. Results.**

High-spatial-resolution simulations show RDF during the initial release of IR-active particles in a nested version of MarsWRF. Fig. 1 shows results for release from the midlatitude site for 60 nm diameter Al rods at a rate of 60 L/s. The plume self-lofts and thickens the Planetary Boundary Layer (PBL) height (Fig. 1a,b show column visible opacity and PBL height at 11AM local time at the release site), which is ~2× thicker than without self-lofting (Fig. S7). Above the release point (Fig. 1c, top row), particles accumulate in the nighttime PBL before mixing upward during daytime. In adjacent grid boxes (Fig. 1c, middle row), particles mix across the surface layer during late night and predawn, while early evening winds leave this region particle-free. Several grid points downwind (Fig. 1c, bottom row), a detached high "anvil" of particles dominates the nanoparticle mass mixing ratio (mmr) between 5 and 10 km above the surface. Further from the plume (Fig S7), the particle mmr peaks between about 7.5 and 12.5 km, at or above the top of the plume-free PBL. The vertical distribution of particles evolves as follows. While initially depleted at altitude, within 100 days an aerosol-enriched layer forms above 7 km. During southern spring, the strong Hadley cell temporarily halts particle buildup (Fig. S6). These vertical variations diminish to 10% over years as mixing homogenizes the atmosphere. The steady-state particle distribution shows only a modest longitudinal enhancement (15° full-width-at-half-maximum) centered on the release site. Particle optical depth in steady state varies seasonally by 15-20%, rising during northern summer when winds are weakest, regardless of release location.

Fig. 2a-f shows the global spread of nanoparticle mmr in lower resolution simulations (5°×5°), for release at an equatorial site. The lower resolution is used to allow many more simulations to be undertaken than is possible for nested cases (such as for Fig. 1). The high resolution allows more detail to be seen in the plume distribution but does not change the main results. The warming plume spreads eastward at all altitudes and also westward at high altitude in the high latitudes, with inter-hemispheric mixing occurring within months. These particles achieve warming much more efficiently than previous designs—a visible optical depth ($\tau_{vis}$) of 0.2 requires only ~45 mg/m$^2$ of aluminum rods in the aluminum simulations, and the same optical depth requires just over ~17.5 mg/m$^2$ of graphene in the graphene simulations. The particles are more than twice as radiatively effective (in the thermal IR, per unit mass) than previous designs. The warming atmosphere inflates both in the sense of vertical expansion due directly to increased scale height, but also due to reduction in total mass of the seasonal $CO_2$ ice caps, see Table S3. This inflation combined with the RDF self-lofting lifts the suspended particles high into the atmosphere. The system approaches steady state with an e-folding timescale of ~1.1 Mars years (Fig. 3, Fig. S8). The total time to achieve warming is about the same as the particle atmospheric lifetime. Cooling after particle release ends follows the same timescale. This timescale emerges from the model as the timescale for mixing any aerosol with similar radiative properties. Nothing has been done to make the particles mix more rapidly.

Fig. 2d shows warm season surface temperatures. The warm season surface temperature is defined as the average surface temperature during the warmest 36° of solar longitude (~70 days) of the year. This corresponds to southern summer in the southern ground ice zone, and northern summer in the northern ground ice zone.

The warmed climate develops a stronger meridional overturning circulation (Fig. 4). The Hadley cells strengthen by a factor of 4 and shift upwards while the seasonal strength asymmetry (Richardson & Wilson 2002) largely disappears. Because the particles are relatively well mixed in steady state, the mass streamfunction corresponds to the meridional overturning circulation of particles. Near-surface winds increase by 60% (global and annual average). Steeper temperature gradients develop between sunlit regions and the winter pole. This is the opposite of Earth's $CO_2$-warming response, which preferentially warms the poles and weakens circulation (Xia et al. 2020).

Surface pressure increases as seasonal $CO_2$ seasonal ice caps shrink. We do not consider release of additional $CO_2$ from perennial $CO_2$ ice nor from adsorbed regolith reservoirs, which would at least double the atmospheric pressure under warming (Bierson et al. 2016, Buhler & Piqueux 2020). This thicker $CO_2$ atmosphere would add to the greenhouse effect, moderate the equator-pole temperature gradient, and expand the percentage of surface area above the triple point of water (Dickson et al. 2023).

The IR-active aerosols spread broadly, unlike patchy dust storms on Earth and Mars, because they settle very slowly. When artificially ballasted to give a settling speed corresponding to the diameter of natural Mars dust (3 μm), the particles produce strong concentration gradients—column depth varies by 5× between the release latitude and south pole, and particles remain below 200 Pa altitude. This creates localized warming, with a 2000 km-wide +10 K warming zone, though less efficiently than global warming.

Among the two release sites tested, steady-state temperature is modestly warmer for equatorial release (Fig. S9), with greater temperature differences for higher particle loads. The steady-state particle loading is also slightly higher, as expected given the longer distances (and correspondingly longer time for sedimentation) for particles released from a far northern site to spread globally.

**4. Discussion.**

For the first time, we modeled the atmospheric dynamics of IR-active particles released from Mars' surface in a global 3D model. Further research is needed on both Mars and warming options before committing to any plan for Mars' future (DeBenedictis et al. 2025). For example, the case for maintaining Mars' surface as a pristine wilderness indefinitely is given in Marshall (1993). Moreover, making Mars' surface suitable for life involves many steps beyond initial warming, for example, soil chemistry and suitability for biology (DeBenedictis et al. 2025). Progress here would be useful for both investigations of Mars' possible future habitability and for astrobiology research (Wordsworth & Cockell 2024, Wordsworth et al. 2025).

As warmer climates have more water vapor, and water vapor is a greenhouse gas, water vapor feedback will strengthen warming. Water ice clouds have also been proposed to provide strong positive feedback for Mars warming (*e.g.*, Wilson et al, 2008, Madeleine et al. 2014), although this depends on grain size and cloud height. As the water cycle intensifies, aerosol may act as ice nuclei / cloud condensation nuclei, which may scavenge aerosol. Even if snow does not reach the ground, scavenging can still transport aerosol vertically downward prior to sublimation (i.e., virga). These examples show that water cycle feedbacks require further study. Additionally, a more active water cycle within a circulation that has enhanced “dust storm-like” global circulation (Fig.4d) will likely greatly increase water atmospheric loss rates to space (Heavens et al. 2018, Chaffin et al. 2021) above the current rate of approximately 0.005% of Mars' confirmed water inventory per million years.

Several unmodeled dust feedbacks are potentially important. Initially, stronger surface winds would lift more dust (Hartwick et al. 2022), creating a potential negative feedback since dust reduces dayside temperatures despite causing net warming (Streeter et al. 2020). However, strong warming would eliminate the $CO_2$ ice cap edge where strong dust lifting generates most Martian dust storms (Kahre et al. 2017). Additionally, dust can be trapped both by surface liquid water and in high-altitude 'duststones' (Bridges & Muhs 2012).

Aerosol-aerosol and aerosol-surface interactions warrant further research. Agglomeration of particles (Bertrand et al. 2022) is most likely during dispersal. Scavenging by condensation of water onto the engineered aerosol could remove a significant fraction of the particles. Possible mitigations include anti-stick coatings, more dispersal points, charging, and day-only release. Since coatings can be very thin (atomic scale), they need not matter much for radiative properties. Agglomerates are deliberately produced during aerosol synthesis (and their properties tuned) (Friedlander 2000, Buesser & Pratsinis 2012), and it is possible that agglomerates will remain both efficient at Mars warming and also easily wind-transported, depending on their fractal dimension. Agglomeration between engineered aerosols and natural dust might also occur. Few data exist regarding dry deposition of submicron particles on desert surfaces, and this uncertainty is important (e.g., Zhang & Shao 2014, Emerson et al. 2020, Li et al. 2025). If the dry deposition rate is 3× greater, then the particle lifetime will be 3× shorter for slow-settling particles such as graphene disks.

Particles must be engineered to break down in the natural environment. Methods include adding spacers, designing for water solubility, further thinning for oxidation frangibility, and functionalization. Graphene particles could offer additional benefits if doped with soil nutrients.

In our simulations, even though Mars' low-latitude temperature swings are reduced, they remain ~100 K, and winter-minimum temperatures in the ground-ice-rich latitudes remain cold (<230 K for latitudes poleward of 30°). These swings will moderate as Mars' atmospheric thickness further increases but will remain severe by Earth standards.

Other particle release strategies might lead to stronger warming than the continuous and steady one used here. For example, particles could be released during the season when the circulation at the site is upward, or at locations and times that have strong diurnal updrafts.

We note that although self-lofting helps particles rise and spread for initial deployment (Fig. 1 and Fig. S7), and we run to steady state, we have not determined whether the self-lofting is important in maintaining, as opposed to establishing, the steady-state warmed atmosphere.

Either carbon or metal particles could warm Mars from near-surface release, if produced at scale. The details of how particles are prepared with both high throughput and sufficient uniformity are challenging (Supporting Information).

Analyzing Mars with stronger longwave forcing but the same orbital forcing is scientifically interesting, especially as comparison to Mars' geologically recent past, with similar longwave forcing but different orbital forcing, is a major scientific goal (MEPAG 2025).

## 5. Conclusions.

Using a particle-tracking climate model, we simulated the atmospheric dynamics of warming Mars with engineered aerosol. We find that radiative-dynamical feedbacks, including local self-lofting and stronger large-scale circulation, aids particle spread. This study addresses only some aspects of the question of IR-active particle release might modify Mars' climate: atmospheric processes are inherently complex, and many open questions remain. These include water cycle feedbacks and agglomeration mitigation approaches.

**Acknowledgements.** We thank C. Willard, A.S. Braude, M. Wang, M. Hersam, R. Zubrin, N. Myrhvold, D. Keith, C. Lee, A.P. Raman, M. Mester, and the PlanetWRF development team. This work used GrADS (COLA/IGES). This work used resources from the University of Chicago's Research Computing Center. Part of the resources supporting this work were provided by the NASA High-End Computing (HEC) Program through the NASA Advanced Supercomputing (NAS) Division at Ames Research Center. This research was supported in part through the computational resources and staff contributions provided for the Quest high performance computing facility at Northwestern University which is jointly supported by the Office of the Provost, the Office for Research, and Northwestern University Information Technology. A portion of this work was conducted at the Jet Propulsion Laboratory, California Institute of Technology, under a contract with NASA.

**Open research / data availability statement.** Model output corresponding to the figures shown is available, see Richardson & Kite (2025). FDTD: 3D Electromagnetic Simulator is commercial code (Lumerical). The MarsWRF source code can be made available by Aeolis Research pending scientific review and a completed Rules of the Road agreement. Requests for the MarsWRF source code should be submitted to mir@aeolisresearch.com.

**Conflict of interest statement.** The authors declare there are no conflicts of interest for this manuscript.

**References.**

**Supporting References.**

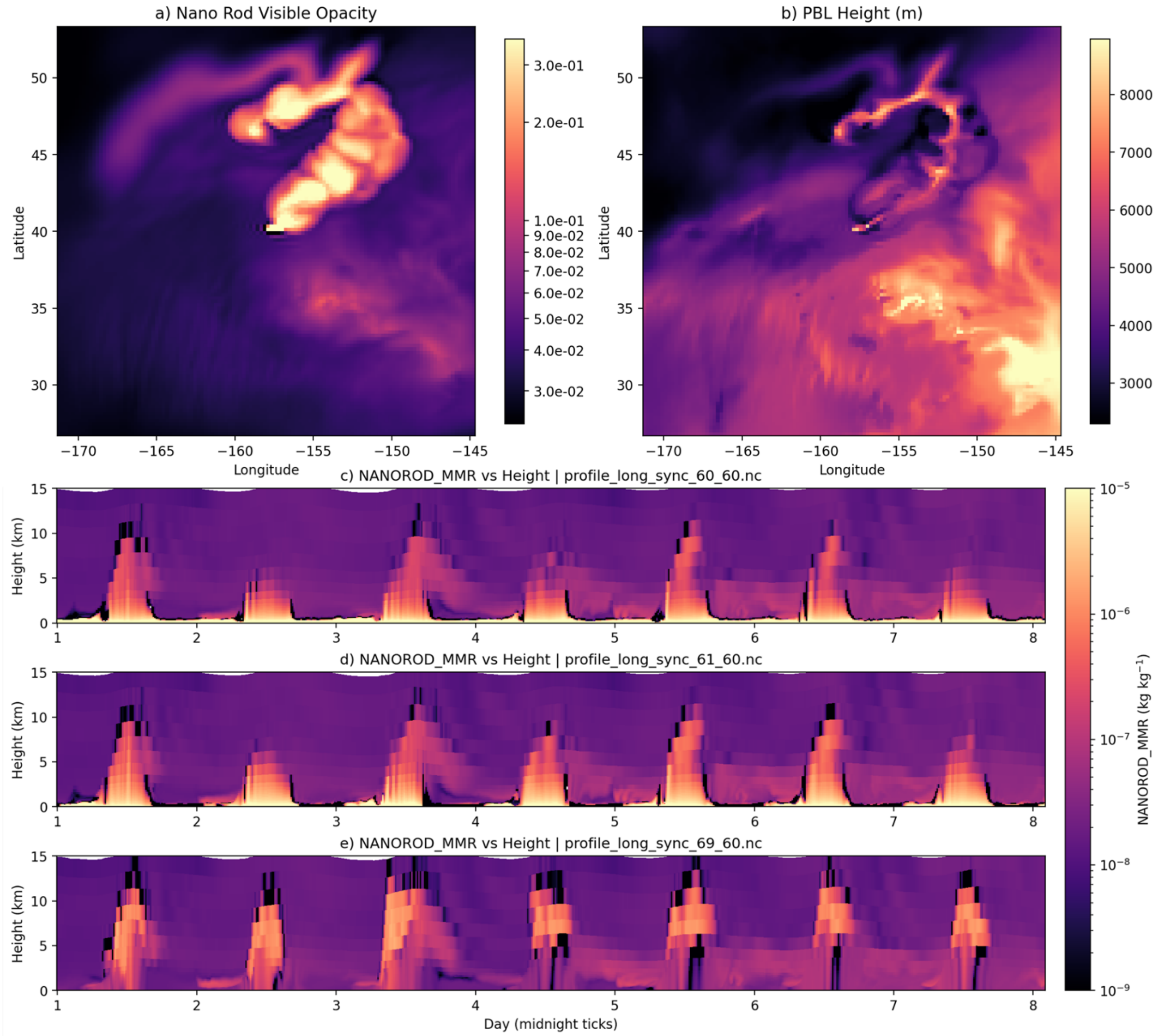

**Fig. 1. Local plume dynamics during initial deployment of 60 nm diameter Al rods at 60 L/s.** Results are captured ~6 sols from the time the plume is initiated. **(a)** Column-integrated opacity of particle plume at wavelength = 0.67 µm ($\tau_{vis}$, unitless). **(b)** Planetary boundary layer (PBL) height (m). Both panels are for the same timestep, with a true local solar time of 11AM at the source site. A mix of shadowing of the ground by plume particles and radiative heating of plume particles leads to a mixture of PBL suppression and augmentation in different locations. **(c-e)** Time-height cross-sections of plume mass mixing ratio in the lowest 10 km. **(c)** is centered over the plume release site, **(d)** is one grid point (~10 km) to the east, and **(e)** is ~100 km further east. Nighttime accumulation in the stable surface layer is evident as a lighter color / yellow band; daytime convection ventilates the accumulated particles deep into the atmosphere, which can then advect downwind in the free atmosphere. Results are for release at Arcadia Planitia (202°E 40°N). MarsWRF has been set up to nest from 2°×2° GCM domain, with two levels of nesting. The nested domain shown has 120×120 grid points and a grid spacing of 0.222°, corresponding to less than 13 km.

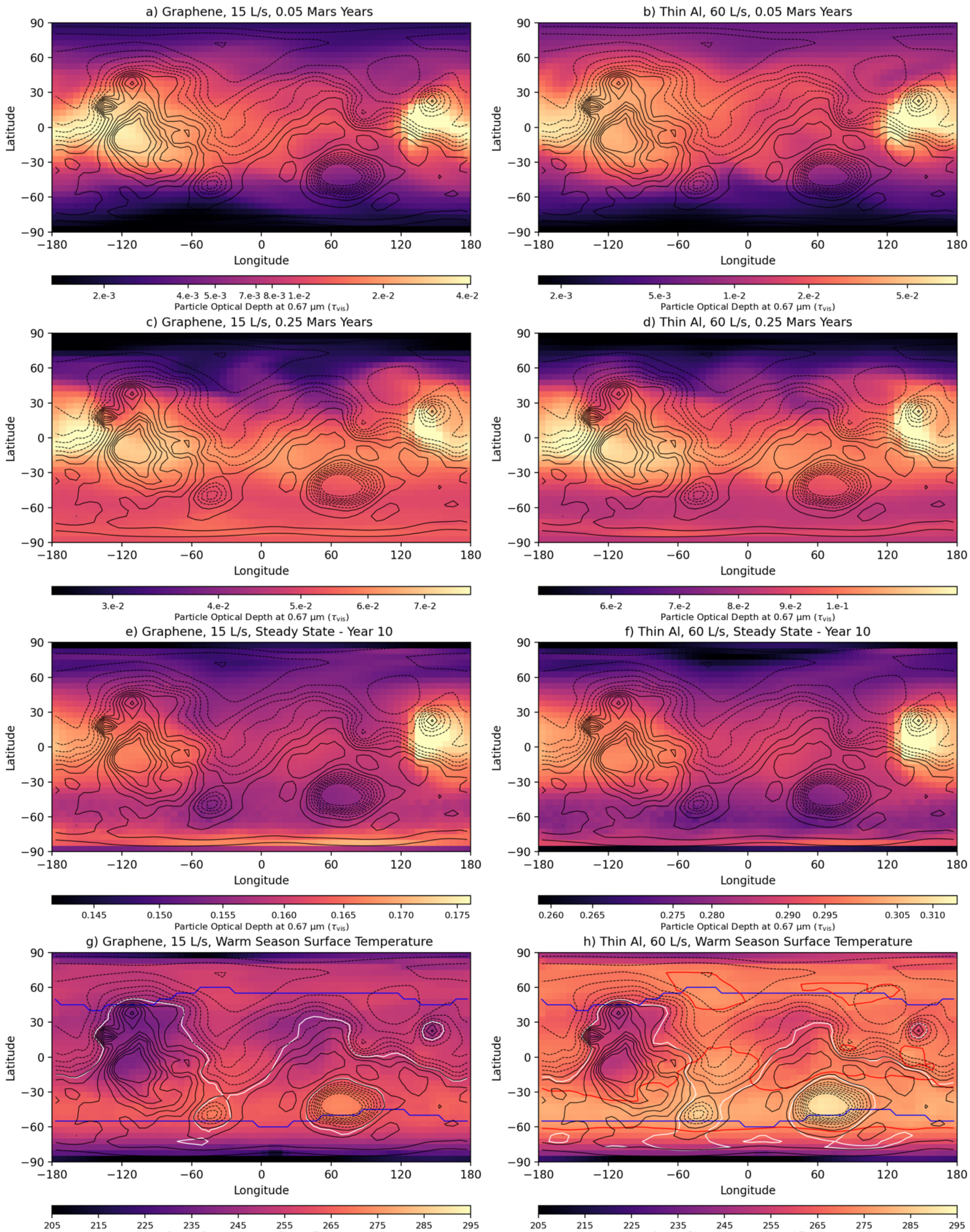
a) Graphene, 15 L/s, 0.05 Mars Years
b) Thin Al, 60 L/s, 0.05 Mars Years
c) Graphene, 15 L/s, 0.25 Mars Years
d) Thin Al, 60 L/s, 0.25 Mars Years
e) Graphene, 15 L/s, Steady State - Year 10
f) Thin Al, 60 L/s, Steady State - Year 10
g) Graphene, 15 L/s, Warm Season Surface Temperature
h) Thin Al, 60 L/s, Warm Season Surface Temperature
Latitude
Longitude
Particle Optical Depth at 0.67 μm ($\tau_{vis}$)
Warm Season Diurnal Average Ground Temperature (K)

**Fig. 2. Dynamics of particle spread and steady-state global warming.** Left: 15 L/s carbon (graphene) disks (*run Cc41*). Right: 60 L/s metal rods (*run Cc16*). Both assuming 0°N 135°E release. **(a-f)** Particle optical depth at 0.67 µm ($\tau_{vis}$). **(g and h)** Filled color: warm season temperatures (K). Black topographic contours correspond to elevations of −5 and −2 km (dashed), and 0, +2, and +5 km (solid). White contour: 610 Pa (~6 mbar) mean pressure level. Blue contour: Approximate equatorward extent of $H_2O$ ice at <1 m depth based on GRS data (Feldman *et al*., 2004). Red contour: Highlights warm-season average surface temperatures above 273K.

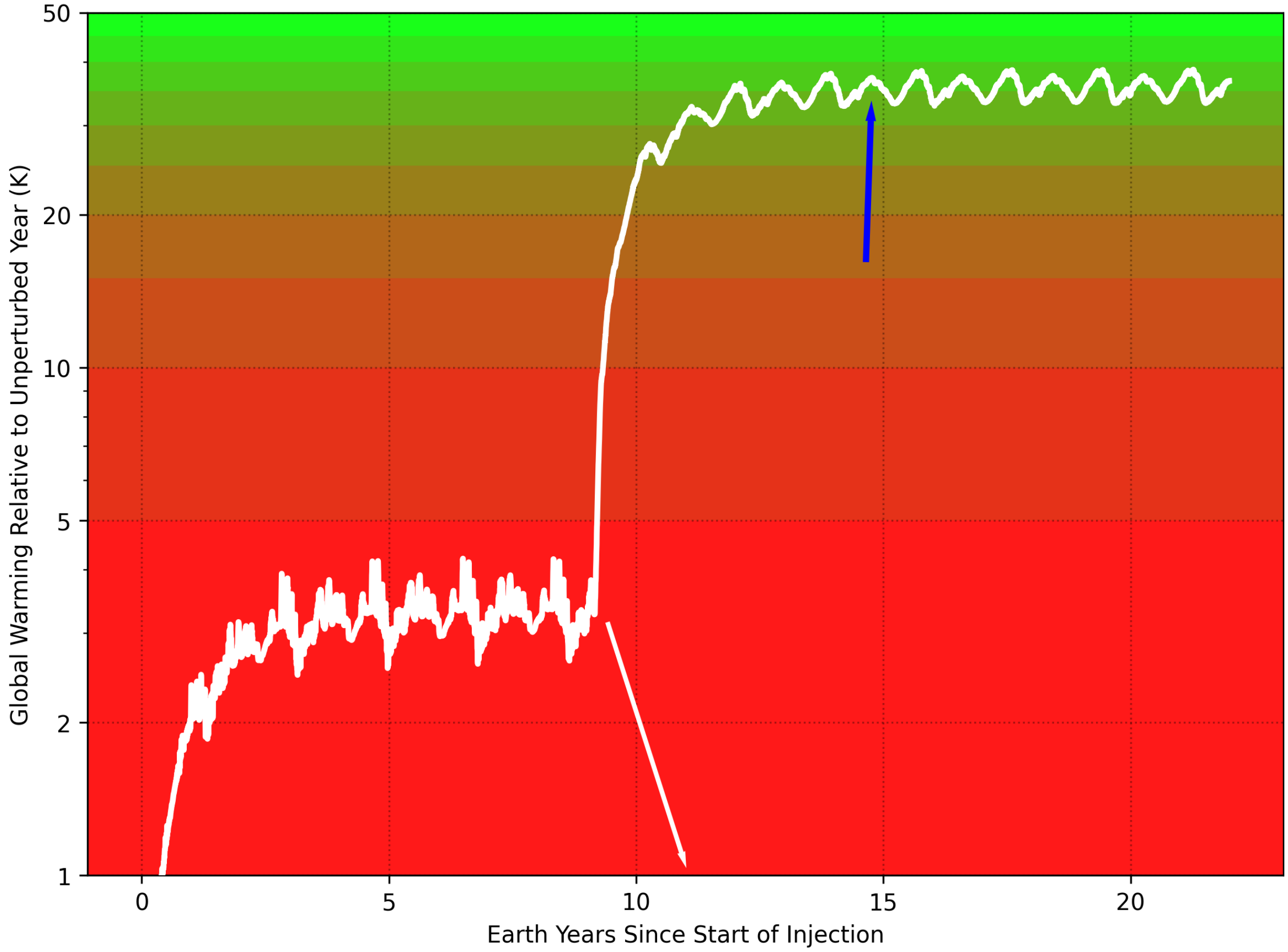

**Fig. 3.** Time evolution of global average surface temperature, assuming a 3 L/s IR-active particle release starting at the northern equinox for the first 5 Mars years (just over 9 Earth years), followed by increase to 60 L/s. Al particles assumed. These two different rates show that the response timescale is almost independent of the release rate. After 8 Mars years (15 Earth-yr, indicated by blue arrow), the average warm-season temperature at 47.5°S exceeds 280 K. The line is constructed by taking the daily global average surface temperature and removing from it the global average temperature for the same day-of-year (sol) from the steady-state reference run. Note that net warming is a modest function of season due to the distribution of nanoparticles. If a choice is made to cease warming, then the planet rapidly cools to the no-warming state (roughly following the white arrow to zero warming).

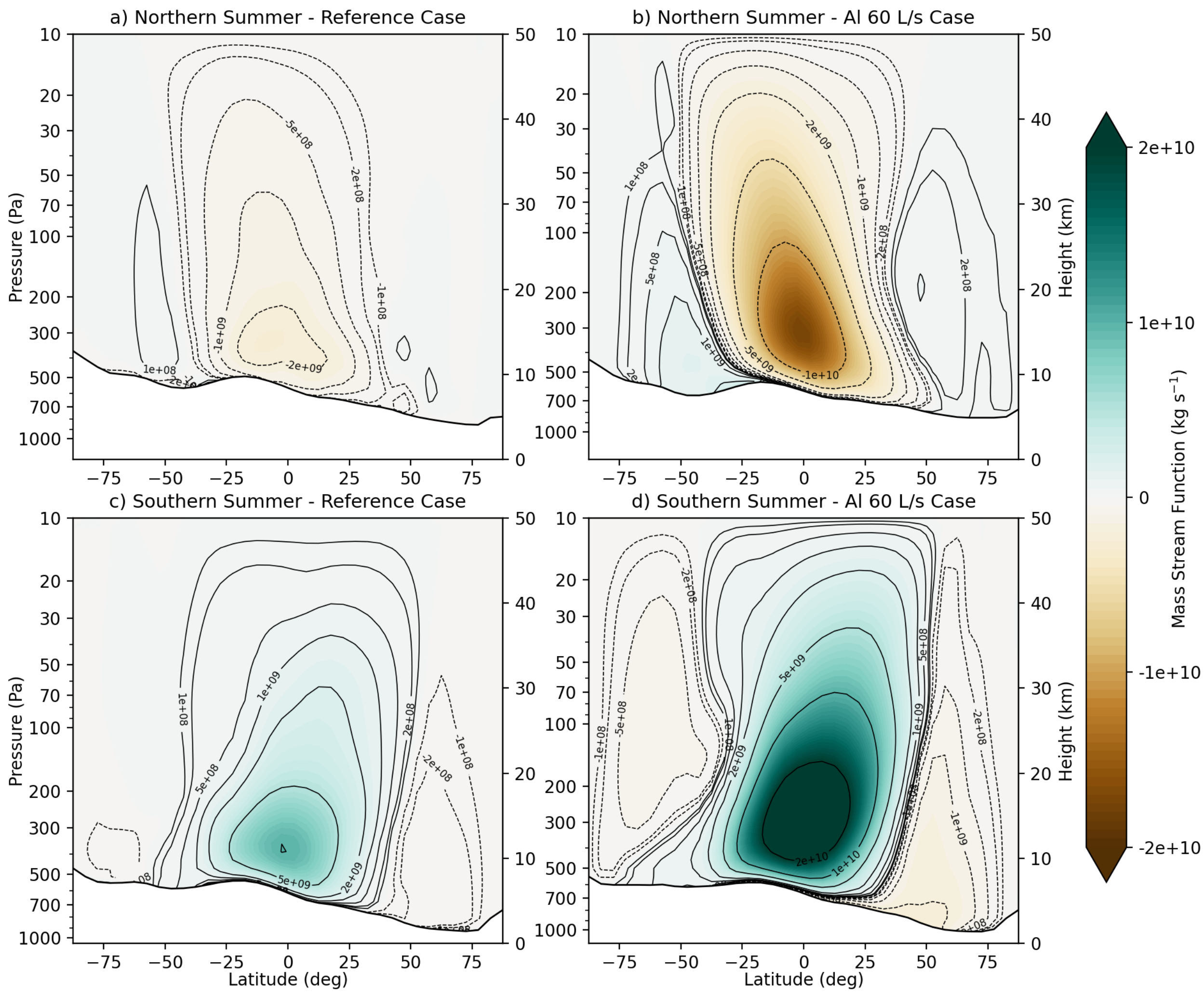

**Fig. 4. Atmospheric circulation strengthens under aerosol warming.** Streamfunctions (kg/s) for **(a, c)** control and **(b, d)** 60 L/s Al-particle warming cases. (a-b) Northern summer, (c-d) southern summer. Positive values: clockwise mass flux (looking toward the west). The surface pressure changes due to the reduction in seasonal $CO_2$ ice in the warmed case. Note that the plot is limited to 50 km to focus on the lower-to-mid tropospheric circulation, but the model top is at roughly 100 km.

Supporting Information for

# Atmospheric dynamics of IR-active particles released from Mars' surface.

Mark I. Richardson[1,*], Samaneh Ansari[2], Bowen Fan[3,4], Ramses Ramirez[5], Hooman Mohseni[2], Michael A. Mischna[6], Michael H. Hecht[7], Liam J. Steele[3,8], Felix Sharipov[9], Edwin S. Kite[3,*]

1. Aeolis Research, Chandler, AZ.

2. Northwestern University, Evanston, IL.

3. University of Chicago, Chicago, IL.

4. Now at: Yale University, New Haven, CT.

5. University of Central Florida, Orlando, FL.

6. Jet Propulsion Laboratory, California Institute of Technology, Pasadena, CA.

7. MIT Haystack Observatory, Westford, MA.

8. Now at: European Center for Medium-Range Weather Forecasts, Reading, UK.

9. Universidade Federal do Paraná, Curitiba, Paraná, Brazil.

*Corresponding authors, mir@aeolisresearch.com and kite@uchicago.edu

## Contents of this file

**Supporting text.**

1. Calculation of optical properties of engineered aerosols.

We calculated optical properties for two material compositions: metal particles (one type of Al rod), and carbon particles (two sizes of graphene disks).

In previous work (Ansari et al. 2024), we simulated 9 µm-long metal nanorods with an aspect ratio of 60:1 using the finite difference time domain (FDTD) method in Ansys Lumerical. In this paper, we carried out similar simulations for 8 µm-long nanorods with a width of 60 nm (133:1 aspect ratio). These particles turn out to be more than twice as mass-effective as the previous design. The FDTD method can numerically calculate an object's response to a pulse of light that contains the desired range of wavelengths. The simulated cross-section is square, but results for circular cross-sections are very similar (Fig. S3 in Ansari et al. 2024). The key output needed for input to the climate simulations are scattering and absorption cross-sections, and the scattering asymmetry (Figs. S1-S2).

*Description of metal-particle models.* We used the aluminum refractive-index data from Rakić (1995) as input to our nanophotonics simulations. We saved 150 output points in the wavelength range 0.24-55 µm with a mesh size of 15 nm. The climate models used here require optical properties of the particles averaged over different orientations. To account for the random orientation of the nanorods in the atmosphere of Mars, we rotated the nanorods in 5 azimuthal and 5 polar angles, resulting in a total of 25 orientations. The simulated angles were {0, 30, 45, 60, 90}°. Then, we interpolated the data of the 25 spatial orientations uniformly in angle in space (over spherical coordinates). Finally, we averaged the interpolated results over the first quadrant of the spherical coordinates to find the angle-averaged results (Table S1). A comparison of the angle averaged extinction cross-section and scattering asymmetry for aluminum 9-µm (60:1) and 8-µm nanorods (133:1) is shown in Fig. S2. The 8-µm particles have an extinction cross-section ~1/4 that of the 9-µm particles, but they are 1/8 of the volume and 1/8 of the mass. Therefore, they are ~2× more mass-effective on a per-unit extinction cross-section basis.

*Description of graphene model.* We used Ansys Lumerical's FDTD method for simulating graphene disks. Two designs were considered: graphene disks of 1 µm diameter (for a peak absorption at ~20 µm wavelength) and graphene disks of 0.25 µm diameter (peak absorption at 10 µm wavelength) (Fig. S3).

Graphene's optical response to light can drastically change based on the measurement substrate, but we need to know the optical properties of free-floating graphene. Hence, finding a set of optical material data that correctly captures graphene in atmosphere is quite challenging. We considered two wavelength regimes. (1) For wavelengths greater than 8 µm, a Drude model (Fermi level energy of 0.6 eV) is used to estimate the conductivity of a monolayer of graphene. The simulations in this regime use a 2D geometry (which doesn't have thickness). The mesh in these simulations is 10 nm. (2) The Drude model is not accurate at shorter wavelengths due to electronic resonance modes in graphene. For wavelengths between 0.2 µm and 8 µm, we used the experimental data of Tikuišis et al. (2023). While the graphene is on a silicon carbide substrate in these experiments, the authors attempted to minimize the interaction of graphene with the substrate in order to make the measurement quasi-free standing. Our simulations in this regime use 3D geometry. The thickness of the 3D cylinders of graphene is set to 0.3 nm to mimic the thickness of a monolayer of graphene. The *z*-direction mesh in these simulations is 0.05 nm around the graphene and 10 nm everywhere else. In all other directions the mesh is 10 nm uniformly.

In total, we ran three simulations to cover the entire wavelength spectrum. First, we use the Drude model to simulate higher wavelength data. To ensure that the simulation times are manageable, we ran two simulations at shorter wavelengths. The first simulation covers wavelengths from 0.2-0.3 µm (which is important because it sets the interaction of the graphene with biologically damaging UV radiation) and the second simulation runs from 0.3-8µm. We concatenated the data of the three simulations to collect 194 data points from 0.2 µm to 55 µm (Table S2).

Since the disks have two axes of symmetry, we rotated graphene disks in the azimuthal angle only. As the 2D geometries cannot be rotated in our simulation tool by default, we assumed a cosine dependency of the simulation output on rotation angle for all wavelengths. When the graphene disk rotates azimuthally, its geometrical cross-section changes by the cosine of the angle. As a result, when the disk is parallel to the light's direction of propagation, the geometrical cross-section is zero. In total, we averaged over 1292 orientations to find the angle averaged outputs. Fig. S3 and Table S2 give the angle averaged extinction cross-section for 1-μm and 0.25-μm diameter disks.

Graphene disks transmit most sunlight but block UV similar to ozone. The graphene UV feature has a center position and full width at half maximum of 0.27 μm and ~0.1 μm, which is similar to that of $O_3$ (Hartley band at 0.255 μm) so has similar UV-protecting effects. Since $\tau_{vis} > 0.2$ (at 0.67 μm) for strong warming (Table S3), and the ratio of graphene absorption at 0.27 μm to that at 0.67 μm is >5, the graphene is optically thick ($\tau > 1$) at the center of the UV absorption feature. Thus, the amount of graphene needed for warming would define an optically thick ($\tau_{UV} > 1$) layer near the boundary between the UV-B and UV-C bands (similar to ozone), reducing the flux of biologically damaging UV at the surface.

*Practicality.* While graphene is a lightweight material with acceptable absorption, some practical concerns need to be answered.

The placement of the absorption peaks of graphene disks changes drastically with varying size of disks, number of layers of graphene, and graphene disks' Fermi energy level. Materials present in the environment of graphene disks can affect their doping level (and hence Fermi energy) causing drastic shifts in the wavelength of the peak of the absorption. Fang et al. (2013) have studied the effect of Fermi energy level and disk size on the graphene's response to light.

Producing a monolayer of graphene and shaping it into uniform disks requires advanced nanofabrication capability and has been heavily researched for nearly two decades (Mbayachi et al. 2021). While some methods of obtaining a monolayer of graphene are more scalable, turning graphene sheets into uniform disks of a certain size is not trivial. Using lithography and etching is a solution that can result in higher uniformity, but at the cost of throughput, making it impractical for warming Mars. Some alternative methods of growing graphene disks have been suggested (e.g., He et al. 2013). However, the throughput and uniformity of these methods might not be enough to produce the type of particles necessary for making climate change. We find that both disk sizes of 250 nm and 1000 nm can warm Mars (Fig. S5), suggesting that a variation in size by a factor of 4 might be acceptable; however, disks resonating at 15 μm would likely not warm Mars much as $CO_2$ already absorbs strongly at that wavelength.

Graphene monolayers can bind via Casimir and van der Waals forces to form multilayers of graphene or graphite (Klimchitskaya & Mostepanenko 2013). This means that the graphene disks tend to agglomerate making storage in small spaces in air impossible. To alleviate the problem, the disks either need to be held in a liquid or be given enough volume that they aren't interacting. This significantly complicates the transportation and dispersion of the particles in Mars' atmosphere.

*Graphene requires modification ("doping") to attain high resonance peak amplitude.* Free electrons in graphene oscillate with the incoming electromagnetic field and create a resonance response in absorption at certain wavelength. As the number of free carriers increase, the resonance peak amplitude increases while the resonance wavelength becomes shorter (Neto et al. 2009). The number of free electrons available in graphene is determined by its Fermi energy. Typically, higher Fermi energies allow for more free carriers and stronger resonance responses. Graphene in its neutral form is a semimetal and does not resonate strongly with electromagnetic field. However, the carrier concentration of graphene can be easily manipulated by using other materials to dope it. Chemical doping is a method of doping graphene (Guo et al. 2011) that is attractive for this application. Graphene can be chemically doped by (1) Substitutional doping where a dopant atom replaces C atoms at some lattice sites. (2) Modulating the doping via chemical species adsorbed on the surface of the graphene (Liu et al. 2010).

However, adsorbed molecules can be desorbed by water or oxygen in the atmosphere. Method (1) is more robust although it introduces new challenges. To do substitutional doping, a dopant precursor must be made available and needs to be broken down so that the atomic form of the dopant is available. These two requirements can significantly change the energy and mass requirement of this solution. As an example, nitrogen is a popular dopant for graphene. Nitrogen is abundant (3% by volume) in Mars' atmosphere. The two main methods of chemical doping with nitrogen are in situ and ex situ (Deokar et al. 2022). In the former there is a need for particular precursors for nitrogen and these need to be present during the growth of the graphene. The latter requires thermally annealing the graphene layer in a suitable environment.

<u>2. How the particles are added to the climate models.</u> The graphene disks are added to the climate model as a 16:1 (number ratio) mix of 250 nm diameter disks and 1000 nm diameter disks. The combined broadside-on area is therefore $16 \times (\pi\ 0.125^2) + (\pi\ 0.5^2) = 1.57\ \mu m^2$. The asymmetry parameter of the mixture, $g_{mix}$, is calculated via

$$g_{mix} = g_p\ (1 - ( r\ ( q_{sca} / p_{sca}) / ( r\ q_{sca} / p_{sca} +1 )) + g_q\ ( r\ q_{sca} / p_{sca}) / ( r\ q_{sca} / p_{sca} + 1 )) \qquad (1)$$

where $r$ is the mixing ratio of $q$ to $p$ in particle number, $q_{sca}$ is the scattering cross-section of particle type $q$, $p_{sca}$ is the scattering cross-section of particle type $p$, $g_q$ is the scattering asymmetry of particle type $q$, and $g_p$ is the scattering asymmetry of particle type $p$.

For numerical reasons, we add the disks to the model in sets of 1700 (i.e., 1600 small disks and 100 larger disks). The particle volume flux was related to the particle optical depth using the equivalent spherical diameter. To calculate the equivalent spherical diameter for graphene, which is a 2D material, we assumed an areal density of 0.77 $mg/m^2$ and a three-dimensional density of 2,267 $kg/m^3$. This gives an equivalent spherical diameter of $2 \times ((0.77\ mg/m^2 \times 1.57\ \mu m^2 / 2{,}267\ kg/m^3) / (4\pi/3))^{1/3} = 0.101\ \mu m$ for 17 disks, and 0.467 μm for a set of 1700 disks. The equivalent spherical diameter for the Al nanorods is $2 \times ((0.06 \times 0.06 \times 8) / (4\pi/3))^{1/3} = 0.38\ \mu m$.

The column mass required for optical depth unity at the reference wavelength of 0.67 μm ($\tau_{vis} = 1$) is calculated as follows. For the graphene disk mix, the orientation-averaged extinction cross-section for a set of 17 disks is 0.0207 $\mu m^2$. The geometric broadside-on cross-section is 1.57 $\mu m^2$. (Graphene is almost transparent in the visible). The areal mass density of graphene is 0.77 $mg/m^2$. Therefore $\tau_{vis} = 1$ is achieved for $(1\ m^2 / 0.0207\ \mu m^2) \times (1.57\ \mu m^2 \times 0.77\ mg/m^2) = 58$ $mg/m^2$ column mass of graphene particles. For the 8 μm-long dimension, 60 nm-short dimension aluminum rods that are considered in this study, the orientation-averaged extinction cross-section of 1 particle is 0.493 $\mu m^2$. The mass of 1 particle is $2700\ kg/m^3 \times 8\ \mu m$ $\times$ 60 nm $\times$ 60 nm = 7.78 $\times$ $10^{-17}$ kg. Thus, the column mass at $\tau_{vis}$ = 1 is (1/0.493 $\mu m^2$) $\times$ 7.78 $\times$ $10^{-17}$ kg = 158 mg/$m^2$ column mass of aluminum particles.

Particles must be designed to degrade; for example, on being brought into human habitats via airlocks, and to prevent unsustainable accumulation/build-up on the Mars surface. Surface albedo changes are not modeled in this study.

3. Calculation of the surface-warming effect of the particles using a 3D climate model. *Details of model setup:* For the MarsWRF global runs to steady state that are shown in Figs. 2-4 and summarized in Table S3, we use a single model grid whose resolution is 60×36×52 latitude-longitude-height. The model top is set at 120 km (the layers are tabulated in Li et al., 2025). For the long-duration global runs, timestep is varied from 30 s to 180 s (the model uses Martian seconds, which are just under 3% longer than SI seconds). The radiative transfer uses a correlated-k scheme, referred to as the "KDM scheme," that treats both shortwave and longwave forcing (Mischna et al. 2012). The KDM scheme uses Planck-weighted averaging (215 K for the infrared bands, 6000 K for the solar bands) to down-sample the high-spectral-resolution engineered-aerosol optical properties to the 14 bins typically used in the KDM scheme (see Ansari et al. 2024 for details). In a very limited number of MarsWRF SCM simulations, we instead use 112 bins. Radiation physics calls are made every 15 Mars minutes. The surface layer scheme is a Monin-Obukhov scheme, and we use a Martian 12-layer subsurface heat diffusion scheme. The planetary boundary layer scheme is a YSU 1st order closure scheme (Hong et al., 2006), with Mars implementation similar to that in Richardson et al. (2007). The water cycle is turned off. The surface thermal inertia map, surface albedo map, obliquity, and orbital parameters are for present-day Mars.

Terrain-following modified-$\eta$ coordinates are used to define the z-levels in the model (where $\eta = P-P_{top}/P_{surf}-P_{top}$, where P is local pressure, $P_{surf}$ is surface pressure, and the fixed $P_{top}$ is the model-top pressure).

Runs are started cold, with an equilibrated atmosphere and $CO_2$ cycle designed to represent modern Mars. They are run for a period exceeding 2 e-foldings of the nanorod loading, with results generally examined after 6 simulated Mars years, though most runs were extended for over a Martian decade. Run output is summarized in Table S3. Runs start at northern spring equinox. Mars' orbit is eccentric, so the time between the northern spring equinox and northern autumn equinox is significantly longer (and less illuminated by the Sun) than the time between southern spring equinox and southern fall equinox. This is why the southern hemisphere warm season is hotter than the northern hemisphere warm season (Fig. S13). Limitations include neglect of dust-air temperature decoupling, which is significant above 40 km (Haberle et al. 2025).

The particle gravitational settling speed (terminal velocity) for rods and disks was calculated using an analytic approach appropriate for the rarefied-gas regime (Sharipov 2016). This is ~4 μm/s for graphene disks at 100 Pa (the settling speed does not depend on the disc size, provided that the size is smaller than the equivalent free path). We impose a dry deposition velocity of no less than 0.3 mm/s in the lowest model layer (~10 m thick), corresponding to non-gravitational-settling particle removal processes (such as diffusion): this is important for graphene. For the Al rods considered in this study, the settling speed at 100 Pa is 0.6 mm/s. MarsWRF internally calculates gravitational settling speed via a Cunningham slip correction formula that assumes spherical particles. Therefore, we set the gravitational settling speed by choosing a particle diameter that gives the correct (analytically calculated) particle fall speed for non-spherical particles. This is 0.073 μm for the Al rods and a negligibly small value (6 nm) for the graphene disk mix. Particle atmospheric lifetime may exceed our estimates since some particles likely get re-lofted, for example, by springtime $CO_2$ sublimation, rather than permanently settling at the surface. We did analytic calculations in the free-molecular limit to determine the diffusivity, D, of graphene disks and Al rods, finding (for p = 1000 Pa and T = 200 K), $D = 6.5 \times 10^{-10}$ $m^2/s$ and $1.0 \times 10^{-8}$ $m^2/s$, for the larger and smaller graphene disks, respectively, and $D = 5.6 \times 10^{-10}$ $m^2/s$ for the Al rods.

*Details of plume-tracking procedure:* For the 'high-resolution' simulations shown in Figure 1, we use MarsWRF as a nested model with two embedded nests. The top-level domain ("d01") has 2×2° resolution (180×90 gridpoints, ~120 km/grid at the equator). We employ the same 52-layer vertical grid as the lower resolution GCM cases. Each level of nesting ("d02" and "d03"), has an increase in resolution of 3× its parent level, providing resolution as high as ~13 km/(grid cell) at the equator in "d03". As illustrated in Figs. 1a, b, the "d03" domain covers ~26.67°×26.67°.

The plume injection sites for these high-resolution runs are the same as the low resolution GCM cases. In Figure 1, only the site situated at 40°N latitude, 202°E longitude is shown. This is the "Arcadia" site in our long duration runs (Table S3). To eliminate start-up transients, "d01" is run for over a Mars year (in actuality, three full Mars years were simulated before the nests were turned on, but this is not necessary) before the nested grids are turned on. Results in Figure 1 are captured ~6 sols from the time the plume is initiated. The lower resolution GCM simulations in Figs. 2-3 employ a coarser resolution of 6×5° (lon × lat), and have no embedded nests, but are otherwise identical in implementation.

The particle mass mixing ratio is always too small to directly affect the atmospheric dynamics via particle weight (in contrast to a terrestrial pyroclastic flow).

4. Calculation of the surface-warming effect of the particles using 1D climate model.

We carried out 2 types of 1D (temperature-versus-altitude) climate modeling: using the model of Ramirez (2017), and using a 1D version of the MarsWRF code.

For the first of these models, the 1D modeling procedure closely follows that used in Ansari et al. (2024) and the following description summarizes and follows previous work. The single-column radiative-convective climate model (RCM) has 201 vertical layers, 55 infrared spectral intervals, and 38 solar spectral intervals. For present Mars (solar flux = 585 W/m$^2$), we assume a typical stratospheric temperature of 155 K and a surface albedo of 0.22 (e.g., Ramirez & Kasting 2017). The 1D model runs warmer (218 K) for the fiducial (no warming) case than does the 3D no-warming case (203.4 K), and the 1D-simulated temperature rises more slowly with increasing particle loading than does the 3D-simulated temperature (reasons are discussed in Ansari et al. 2024). The particles are introduced into the 1D model using the 1D average of the vertical particle mass mixing ratio from a 3D simulation, which is very close to uniform.

The RCM uses log-layers that extend from the ground to the $1 \times 10^{-5}$ bar pressure level (e.g., Ramirez et al. 2017). We employ a standard moist convective adjustment (e.g., Manabe & Wetherald 1967). Should tropospheric radiative lapse rates exceed their moist adiabatic values, the model relaxes to a moist $H_2O$ adiabat at high temperatures or to a moist $CO_2$ adiabat when temperatures are low enough for $CO_2$ to condense. The RCM implements a standard solar spectrum downscaled to model spectral resolution (Thekaekara 1973).

The atmospheric pressure is a Mars-like 650 Pa with an assumed acceleration due to gravity of 3.73 m/s$^2$. Although we prescribe a tropospheric relative humidity of 50%, our results are insensitive to this parameter. The baseline mean surface temperature of the resultant pure $CO_2$ atmosphere (without nanorods) is 218 K, which is warmer than the observed global mean surface temperature of 202 K given the 1D model's lack of clouds, dust, and topography (e.g., Haberle 2013). Nevertheless, the 1D model predicts a similar ~6K difference between the blackbody temperature and the observed one (e.g., Ramirez et al. 2014), indicating that the strength of the natural greenhouse effect is correctly reproduced.

We compute the wavenumber-dependent optical depths ($\tau_v$) for the nanorods with the following (Ramirez & Kasting 2017, Ramirez 2017):

$$\tau_v = 3\, Q_{eff}\, PC\, \Delta z \,/\, (4\, r\, \rho) \quad (2)$$

Here, *PC* is the disk or nanorod particle content (g/m$^3$), $Q_{eff}$ is the extinction cross-section, $\Delta z$ is the path length, $\rho$ is the particle density (kg/m$^3$), and *r* is the equivalent spherical radius (one-half of the diameter calculated in Supporting Text section 2). This equation is then integrated over all disk or nanorod heights and across all wavenumbers.

The particles are well-mixed throughout the atmosphere between the 650 Pa surface and the 10 Pa pressure levels, replacing the 500 Pa surface and 35-km criteria, respectively, assumed in our previous work (Ansari et al. 2024).

As before (Ansari et al. 2024), we assume that the particle content scales linearly with the local pressure. In this work, we compute new RCM mass mixing ratios that yield $\tau = 1$ at a wavelength of 0.67 microns, consistent with assumed GCM optical depth criteria. Thus, instead of keeping PC constant between models, we opt here to maintain a constant optical depth.

We implement a simple procedure to calculate particle warming. For each assumed nanorod optical depth, we find the surface temperature that yields stratospheric energy balance (i.e., the net outgoing and net incoming fluxes must balance each other) (Table S4).

In order to further isolate the differences between the 1D and 3D models, we have additionally run MarsWRF in Single Column Model (SCM) mode. This allows us to eliminate any differences due to physical process treatment (radiative transfer, boundary layer convection, etc.) and numerical choices (grid vertical structure, tropopause fixed temperatures, etc.) and examine only the impact of 1D vs. 3D representation (which is not possible when comparing the RCM with WRF). MarsWRF SCM uses all the same physics as the MarsWRF GCM, but provides a global average treatment. It is forced with the same orbital distance variations (and hence insolation variations) over the course of the annual cycle as the GCM. As the model is much faster than the GCM, we have also experimented with increasing the spectral resolution by a factor of 8. Global average results are shown in Figure S14. These suggest that the SCM with exactly the same physics and spectral resolution as the GCM slightly underpredicts warming for the same total opacity relative to the GCM, and this discrepancy likely results from the error introduced by treating the intrinsically 3D atmosphere as a 1D spatial average. However, holding the SCM model the same except for resolution, we find a larger discrepancy due to spectral resolution, with the low spectral resolution overestimating heating, likely due to unrealistically smearing opacity into wavelength regions that are actually clearer. We note that the 8x spectral resolution WRF SCM is twice as far from the WRF GCM as it is from the RCM (i.e., the non-WRF 1D model) model. Remaining differences are the imposed 155K tropopause temperature and the permanently fixed 585 W/m$^2$ solar forcing. These discrepancies will be examined further in future work.

<u>5. Order-of-magnitude estimates for production of feedstock.</u> Within the framework of our model assumptions, warming Mars by 5 K requires a flux of 5 L/s of Al particles or 2 L/s of graphene disks (Fig. S9). We provide order-of-magnitude estimates of the inputs required.

*a. Graphene.* The chemical energy required to produce graphene from atmospheric $CO_2$ ($\Delta H$ for the reaction $CO_2 \rightarrow C + O_2$) is 393 kJ/mol. From the perspective of production of $O_2$, an essential resource on Mars, this reaction represents a lower energy investment per $O_2$ molecule than the $2CO_2 \rightarrow 2CO + O_2$ process ($\Delta H$ = 566 kJ/mol) demonstrated by the Mars OXygen In-situ resource utilization Experiment (MOXIE) on Perseverance (Hoffman et al. 2022). This is because it recovers some of the energy invested in forming CO. The more energetically favorable reaction has not been considered as a candidate for oxygen production because of the substantial challenge of managing the solid carbon product compared to the relatively simple disposal of CO.

Nonetheless, co-production of graphene and $O_2$ theoretically has the potential to require less overall energy than producing $O_2$ alone. Since a large mass of $O_2$ is needed for propellant for launches from Mars, and humans need ~0.6 tonnes/(Mars year)/person of $O_2$ to breathe, it is of interest to quantify the synergy by coupling graphene and $O_2$ production.

Production of graphene from atmospheric $CO_2$ might be accomplished in two steps: Step 1, like MOXIE, uses solid oxide electrolysis (SOE) to produce $O_2$ and CO by the reaction $2CO_2 \rightarrow 2CO + O_2$. Step 2 converts the CO directly into graphene in a thermally coupled reactor utilizing the exothermic Boudouard reaction $2CO \rightarrow C + CO_2$. The latter step has been demonstrated on a laboratory scale (Grebenko et al. 2022) using chemical vapor deposition (CVD), a technique that lends itself to scale-up and has already been used extensively to produce carbon nanotubes. The power generated by the Step 2 reaction is most readily captured by recycling the hot, pressurized $CO_2$ into the Step 1 SOE system. Note that production of other forms of carbon (e.g., graphite) as an intermediate step to forming graphene could employ the same reaction. Graphene could alternatively be produced from C via methods like liquid-phase shear exfoliation (Paton et al. 2014) or flash Joule heating (Luong et al. 2020).

At a carbon density of $\rho$ = 2,267 kg/m$^3$, target production of 2 L/s graphene corresponds to 382 mol/s carbon (= $O_2$), so 380 ktonnes of $O_2$ /yr. The scale-up of MOXIE is relatively straightforward (Rapp & Hinterman 2023), and it has been estimated that a reference unit producing 838 kmol/yr (3 kg/hr $O_2$) will mass ~1 tonne (Hintermann et al. 2022). The $O_2$ liquefaction hardware would mass another ~1.3 tonnes, for a total of ~2.3 tonnes not including liquefied $O_2$ tanks (part of the system design for launch vehicles and habitats) (Hintermann et al. 2022). Since the gas collection system and other subsystems are common to both SOE and the Boudouard reactor, the latter is likely not to add more than the mass of the SOE itself, or ~0.5 tonne. Power usage for the 838 kmol/yr system is ~25 kW, including liquefaction. The corresponding power system mass depends on what method is used to generate power. A conservative estimate is 6 tonnes (Hintermann et al. 2022), scaling from actual flown-to-Mars solar power gives 1.5 tons (Romero-Guzman et al. 2023), and more optimistic estimates are <1 tonne (e.g., Wollman & Zika 2006). Further scaling the 838 kmol/yr reference system to a system that will produce 2 L/s graphene would thus require ~360 MW. One way to think about scale-up and deployment of a net 2 L/s carbon production is as a network of modular stations each capable of producing appropriate amounts of $O_2$ and power to serve as an exploration basecamp, producing graphene as a secondary product of the $O_2$ generation. Such an arrangement provides flexibility for meeting diverse scientific objectives. As an example, each of 100 distinct sites might be emplaced as exploration infrastructure, with each providing a minimum of 3.6 MW while generating 3.8 ktonne $O_2$ per year. That capability could support, for example, 3 spacecraft launches and an average population of 7,300. The corresponding landed mass per site for the carbon and $O_2$ production is estimated to be ~1253 tonnes for the conservative estimate of power system mass, and ~555 tonnes for the more optimistic estimate of power system mass. Depending on the usage scenario, either additional power would be provided for human life support or else $O_2$ could be stockpiled in tanks and production suspended while the existing power system supported a limited duration mission.

Where measured, organic carbon is only present at low concentration (0.5 kg/m$^3$) in Mars soil (Stern et al. 2022), so we assume graphite is made from Mars air by electrolysis rather than being obtained from soil.

In contrast to $CO_2$ electrolysis, scaling for graphite-to-graphene conversion at the large scales needed to warm Mars is unexplored in a spaceflight context. Scale-up is required to reach the 400 metric ton/day level because the largest graphene producer in the world currently has only 13 metric ton/day capacity (NanoXplore). However, graphene production capacity is increasing rapidly.

*b. Aluminum particles.* We assume 257 MJ/kg to extract Al from soil via molten regolith electrolysis, by analogy with the energy requirement to extract Al from electrolysis on Earth (Table 8 in Staley & Haupin 2000). This first (feedstock) step is likely to be the primary energy sink in the process of producing Al particles. The mining rate of regolith for a subscale (10%) test is ~7 m$^3$/hour at each of 100 sites, assuming 10 wt% $Al_2O_3$ in regolith. Conversion of planetary regolith to metals has been extensively studied in a lunar context (Williams et al. 1979). The major element composition of Mars soil is that of tholeiitic basalt, similar to that of the lunar maria but with more Cl and S. Mg is potentially promising but has not yet been explored.

**Supporting References.**

**Supporting Figures.**

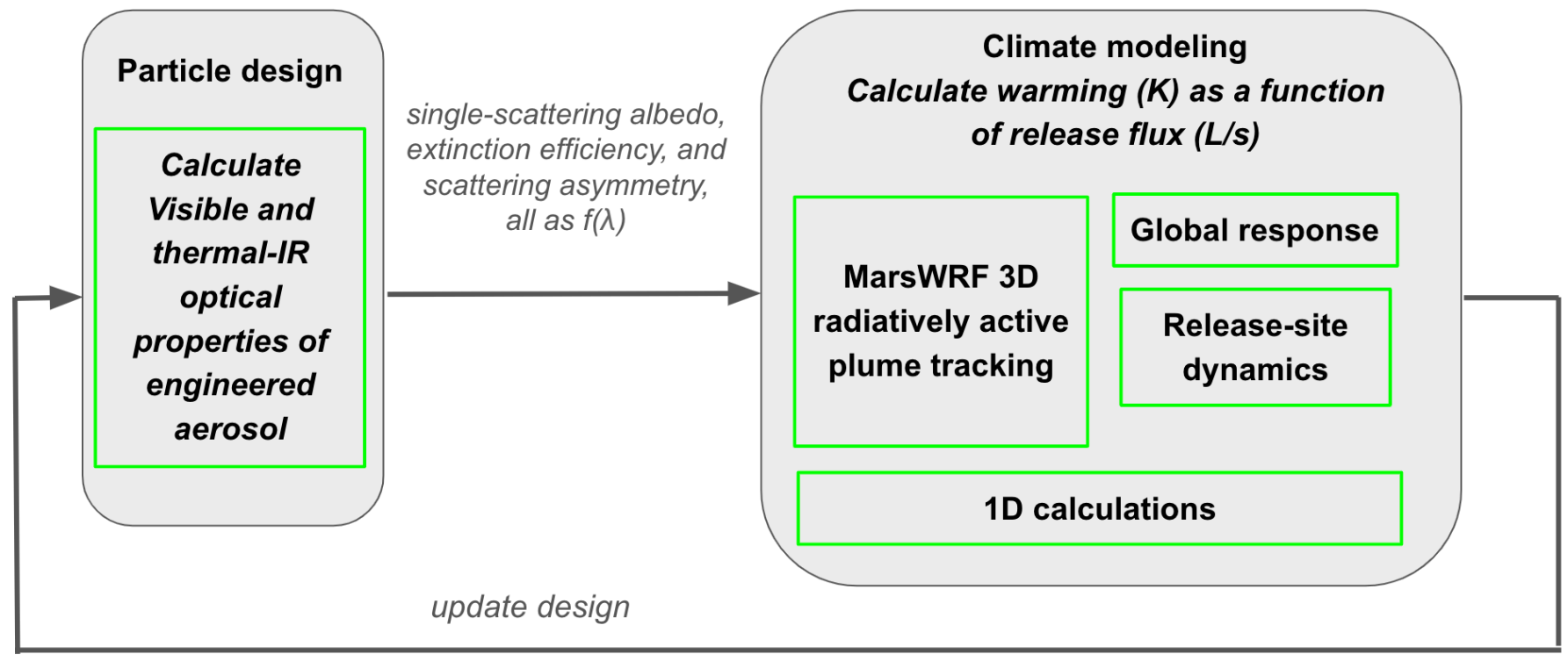

**Fig. S1. Schematic of workflow.**

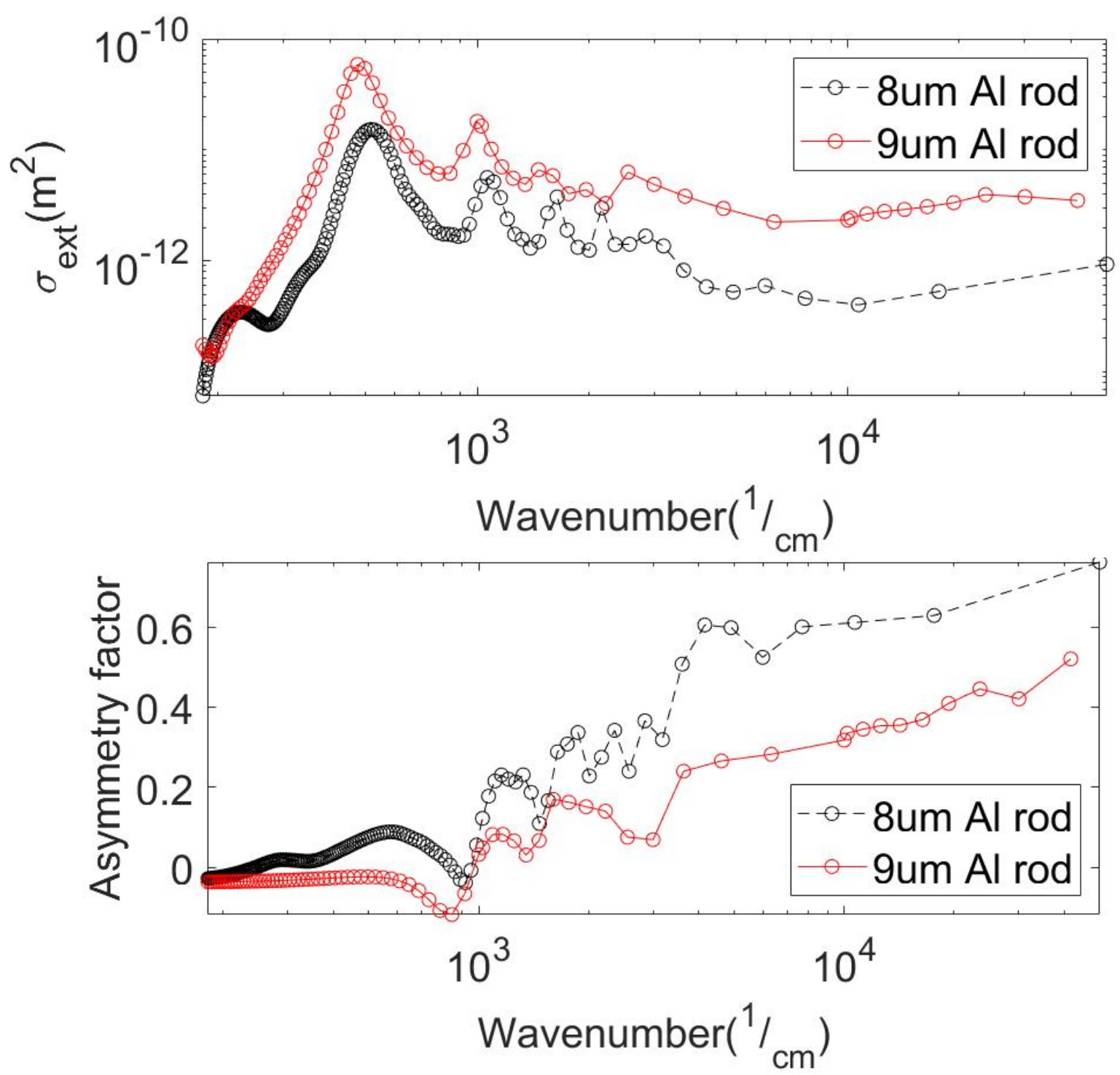

**Fig. S2. Comparison of the angle-averaged scattering asymmetry and extinction cross-section in two Al nanorods.** Shown are the 9-μm (60:1 aspect ratio) design from Ansari et al. (2024) (in red), and the 8-μm (133:1 aspect ratio) design from this paper (in black).

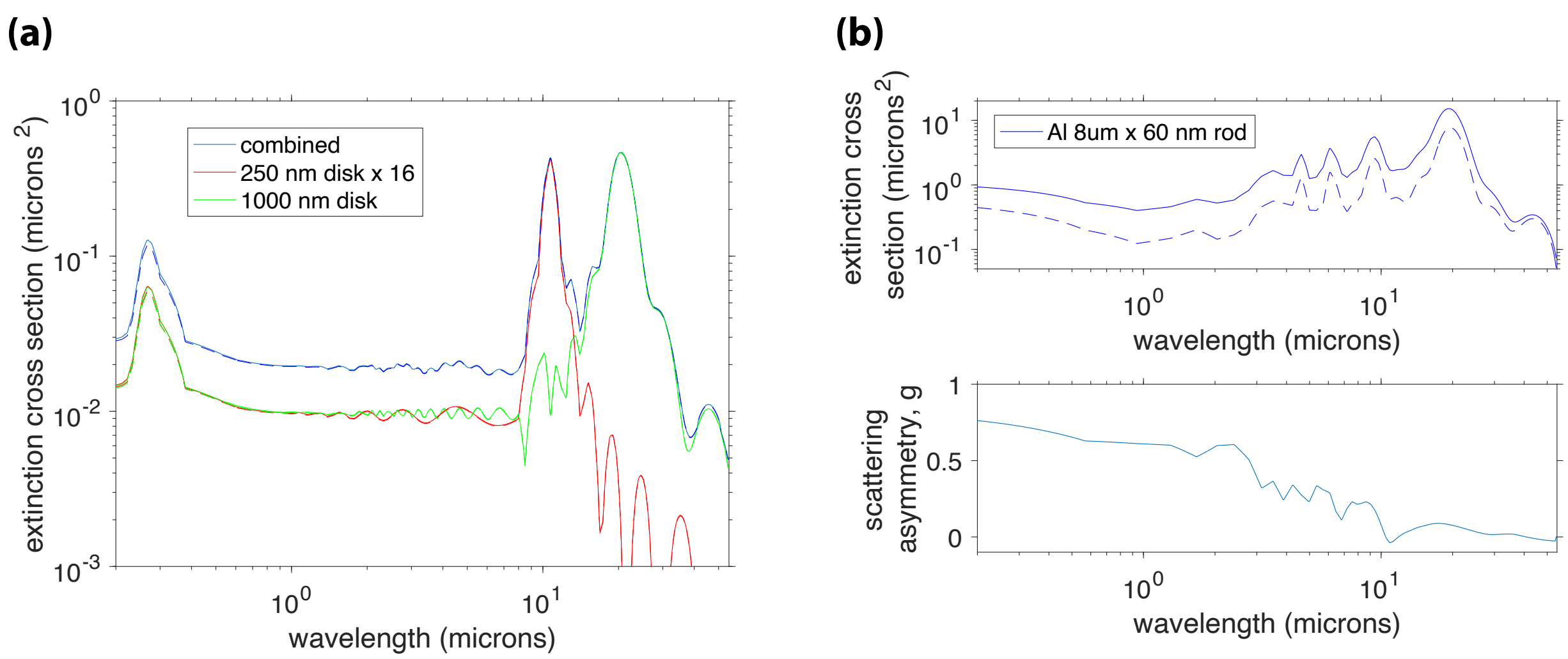

**Fig. S3. Spectra for engineered aerosol** (0.2-55 µm). **(a)** Graphene disk extinction spectrum. Sizes are disk diameters. Dashed lines show the absorption contribution (close to 100% except in the UV). Scattering asymmetry is close to zero. The 250-nm disk is shown multiplied by 16 for ease of comparison, and because a 16:1 number ratio of 250-nm disks to 1000-nm disks is used in most of the graphene simulations. **(b)** 8 µm, 60 nm diameter Al rod extinction spectrum. Dashed line shows the absorption contribution. Lower panel: scattering asymmetry.

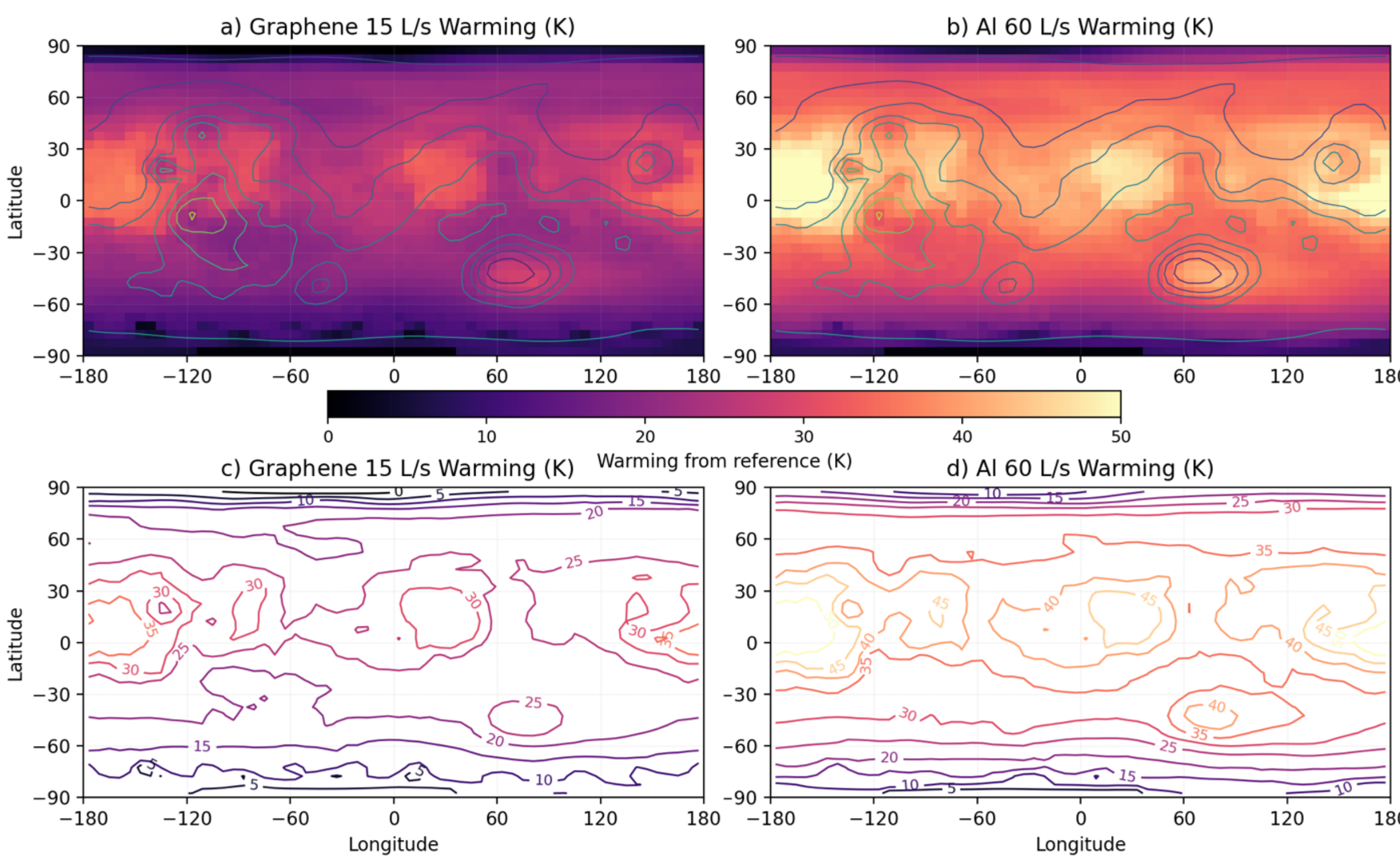

**Fig. S4. Spatial structure of the annual-average warming relative to control case. (a,c)** 15 L/s graphene mix (*run Cc41*). **(b,d)** 60 L/s Al particles (*run Cc16*).

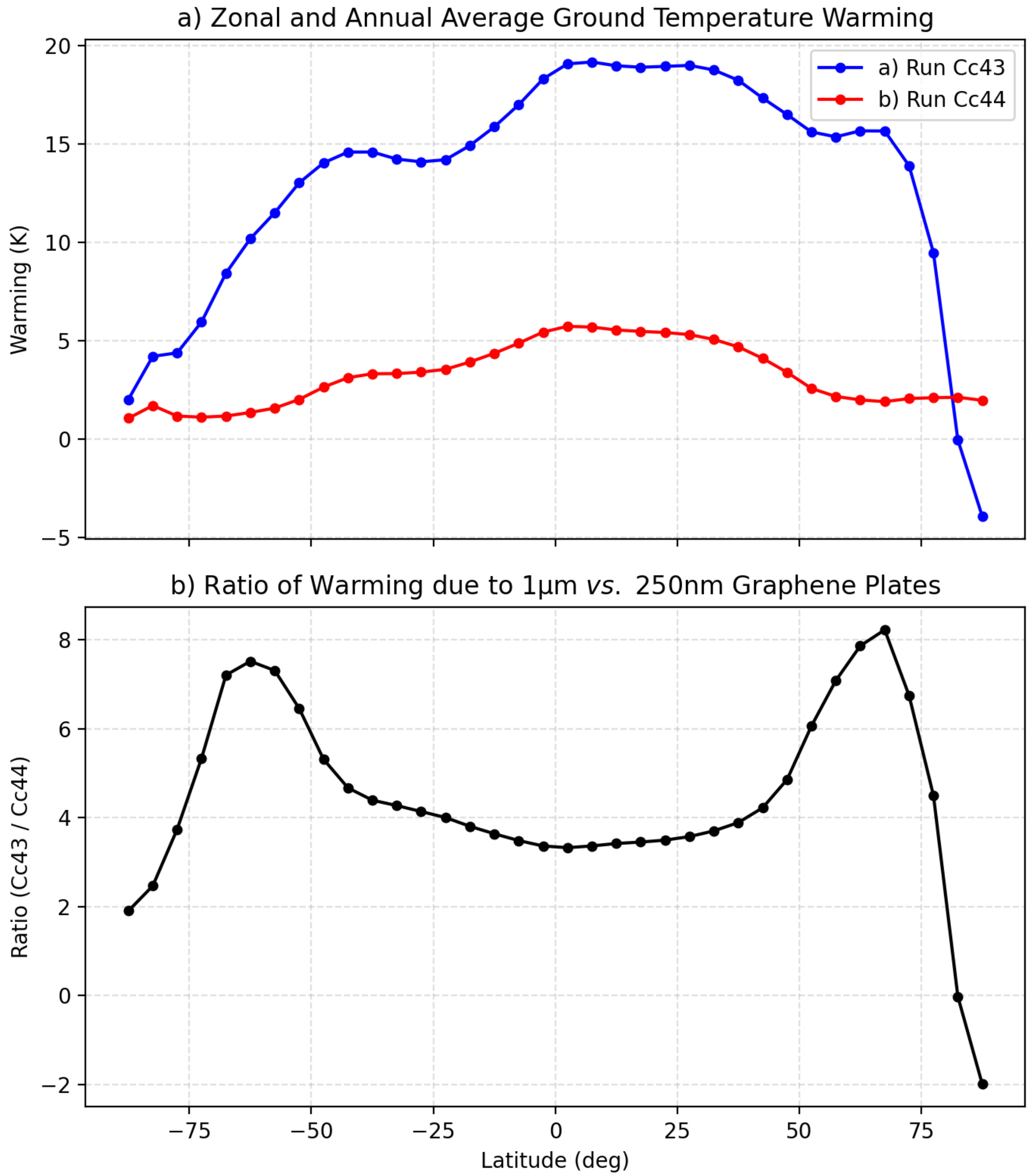

**Fig. S5. Latitudinal tuning of Mars warming. a)** Blue curve shows annual-average warming due to 1 μm diameter graphene disks (black) (*run Cc43*). Red curve shows warming due to 250 nm diameter graphene disks (*run Cc44*). Each curve is generated from the true diurnal average ground temperature calculated on-the-fly from every dynamical timestep (typically 180 s at this resolution) by the GCM. These output quantities are then zonally and annually averaged, and then differenced from the same output generated for the reference (no nanoparticle) case (*run Cc8r*). **b)** The ratio of *Cc43* to *Cc44* warming shows that at the equator,

the 1-µm disks provide just over 3.5x more warming than the 250-nm disks. However, the ratio varies with latitude. Both disk types lose their warming effectiveness at the poles, which remain very cold. At the north pole, the 1-µm disks actually generate net cooling.

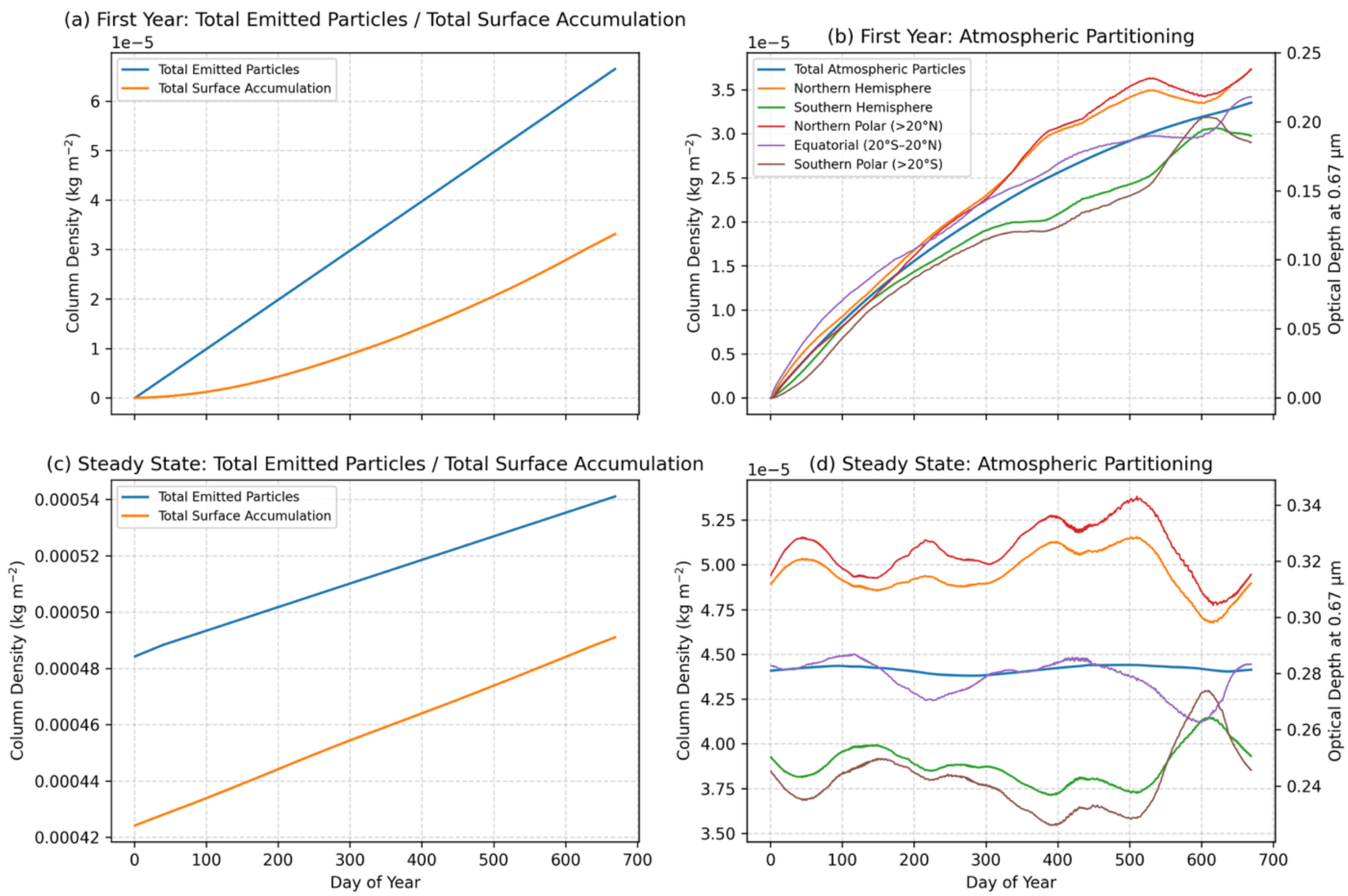

**Fig. S6. Seasonal variation of nanoparticle emission, settling, and atmospheric column density and $\tau_{vis}$. (a,b)** during build-up. **(c,d)** in steady state. For (a,c), blue is emitted particle mass, orange is surface accumulation (both kg/m$^2$). For (b,d) blue = global average, represented as column density on left axis (kg/m$^2$) and 0.67 µm optical depth on right hand axis. The portioning between northern hemisphere (orange) and southern hemisphere (green) is indicated. This global abundance can also be separated into three equal area regions including the equatorial belt (purple), northern "polar" (latitude poleward of 19.47°N, in red), and southern polar (poleward of 19.47°S, in magenta). The x-axis labels the day number of a single Mars year, with day 0 at the northern spring equinox ($L_s$=0°). All output from *run Cc16.*

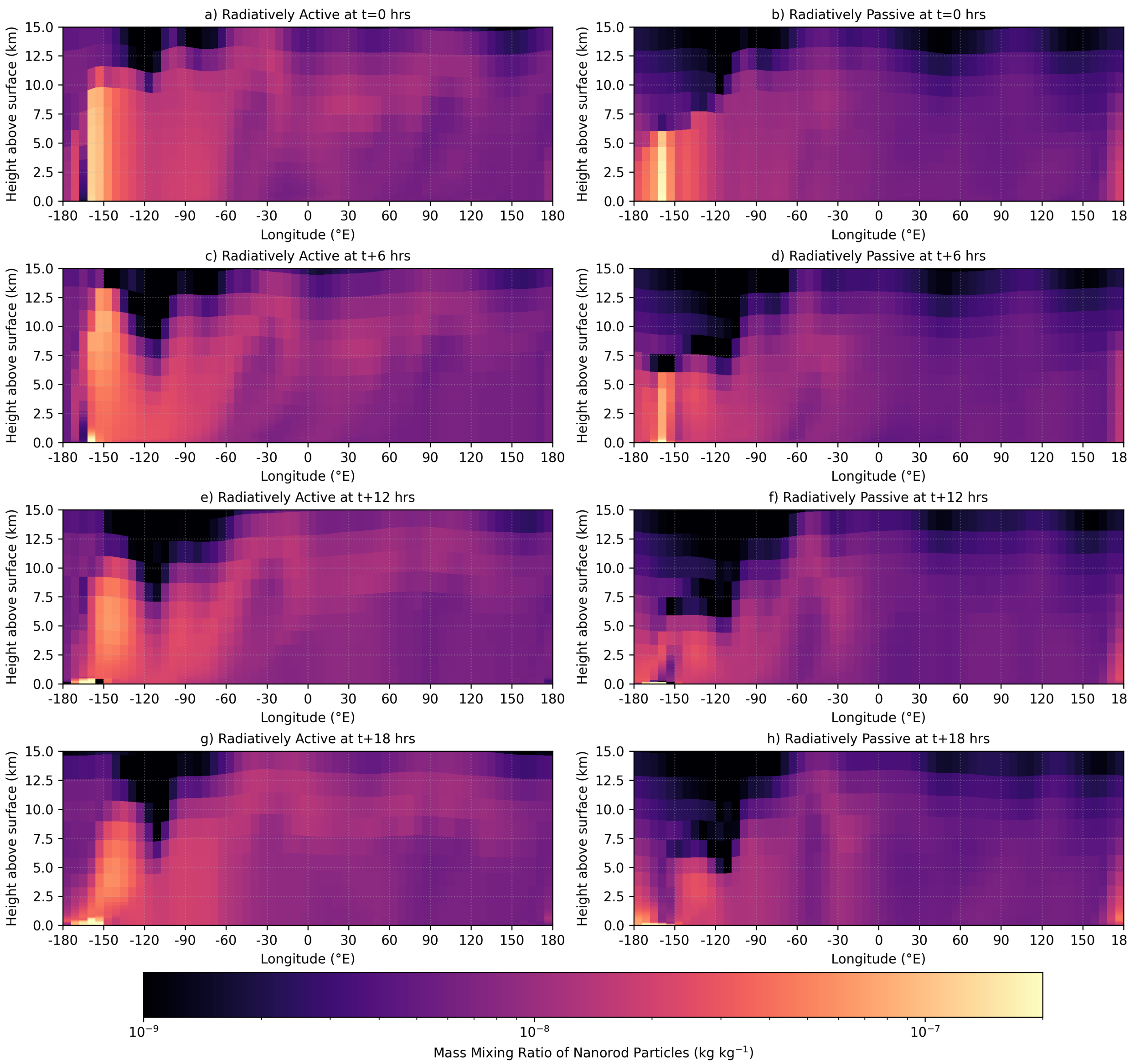

**Fig. S7. Radiative feedback leading to self-lofting and aiding plume deployment.** Longitude-height cross-section through the deployment plume with **(a,c,e,g)** and without **(b,d,f,h)** radiatively active feedback. Shaded quantity is the nanorod mass mixing ratio (dimensionless). (a,c,e,g) 12.5 L/s graphene, ~30 sols into deployment (*run Cc11*). (b,d,f,h) Injected the same 12.5 L/s of graphene but with radiative feedback disabled, ~30 sols into deployment (*run Cc8*). For both simulations, 4 timesteps are shown, 6 Martian hours apart.

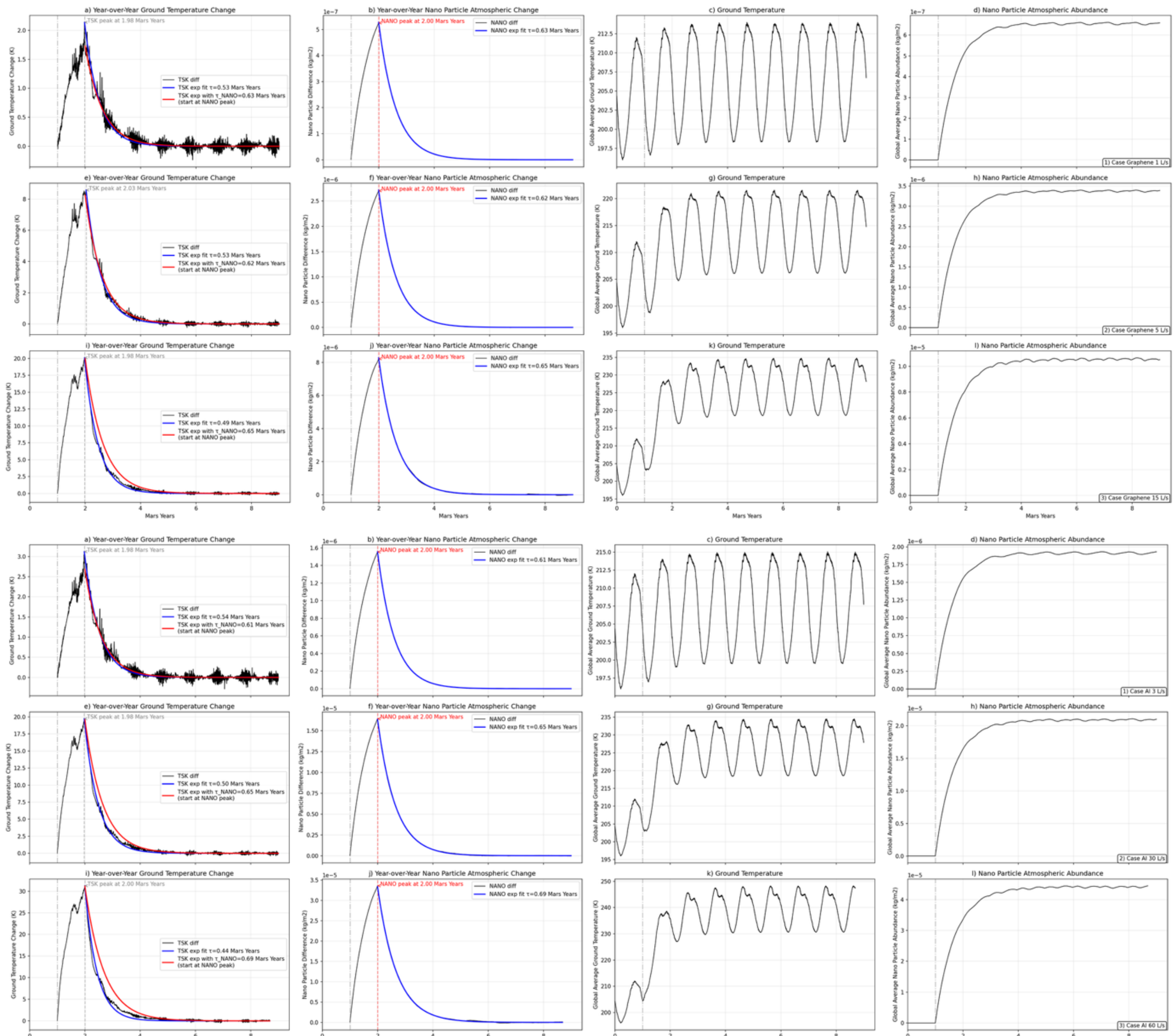

**Fig. S8. Atmospheric nanoparticle increase and ground warming: Initial year increase followed by exponential decline as Mars' atmosphere approaches steady-state warming.** First column: warming relative to the same timestep from the previous year, at each timestep after spin-up (first year excluded as it is in steady state). This algorithm removes the seasonal cycle in temperature. The x-axis units are Mars years. Blue line: fitting the evolution to an exponential decay. $\Delta T$: annual warming. $\tau$: warming timescale. Second column: Year-over-year trends for nanorod abundance (given as average column density). Third column: ground temperature trend. Fourth column: Nanorod abundance trend. All runs shown are for the equatorial release site. Each run is labelled in the lower right corner of the right-most panel.

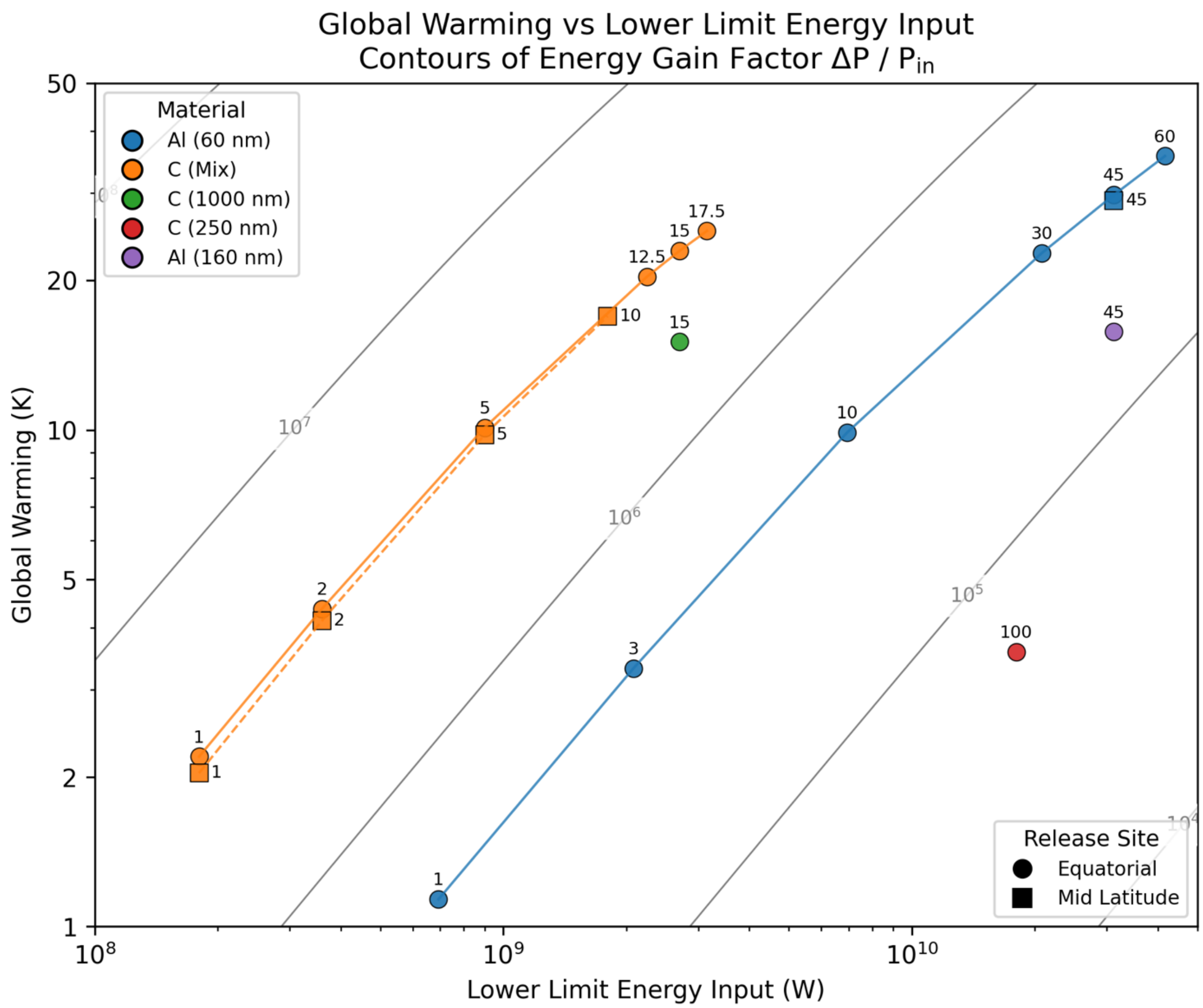

**Fig. S9. Steady-state temperature response as a function of the steady-state energy input.** Data points are from the dynamic plume tracking GCM. The x-axis is the power needed to create the feedstock needed for the specified rate of nanoparticles for injection (see text). Output are shown for all model cases from Table S3 and are as labeled on the plot. The marker numbers correspond to injection rates in L/s. The continuous thin contours show the "energy gain factor", which is calculated as the extra power from the increased ground temperature (the blackbody emission from the new ground temperature - $\sigma T_{av}^4$ - minus the black body emission from the reference case ground temperature) divided by the amount of power needed to make the nanoparticles. The labels on the thin black contours correspond to the numerical value of this gain factor for a given warming and energy input rate. Solid blue line and solid orange lines correspond to release at Elysium. Dashed orange line corresponds to release at Arcadia. The red symbol (*run Cc49*) shows the Al particle design emphasized in previous work (Ansari et al. 2024); the graphene disk mix is ~20× more energy-effective (See Table S3 for detail). Energy input assumes 180 MW/L for graphene disk production and 693.9 MW/L for Al rod production (see text).

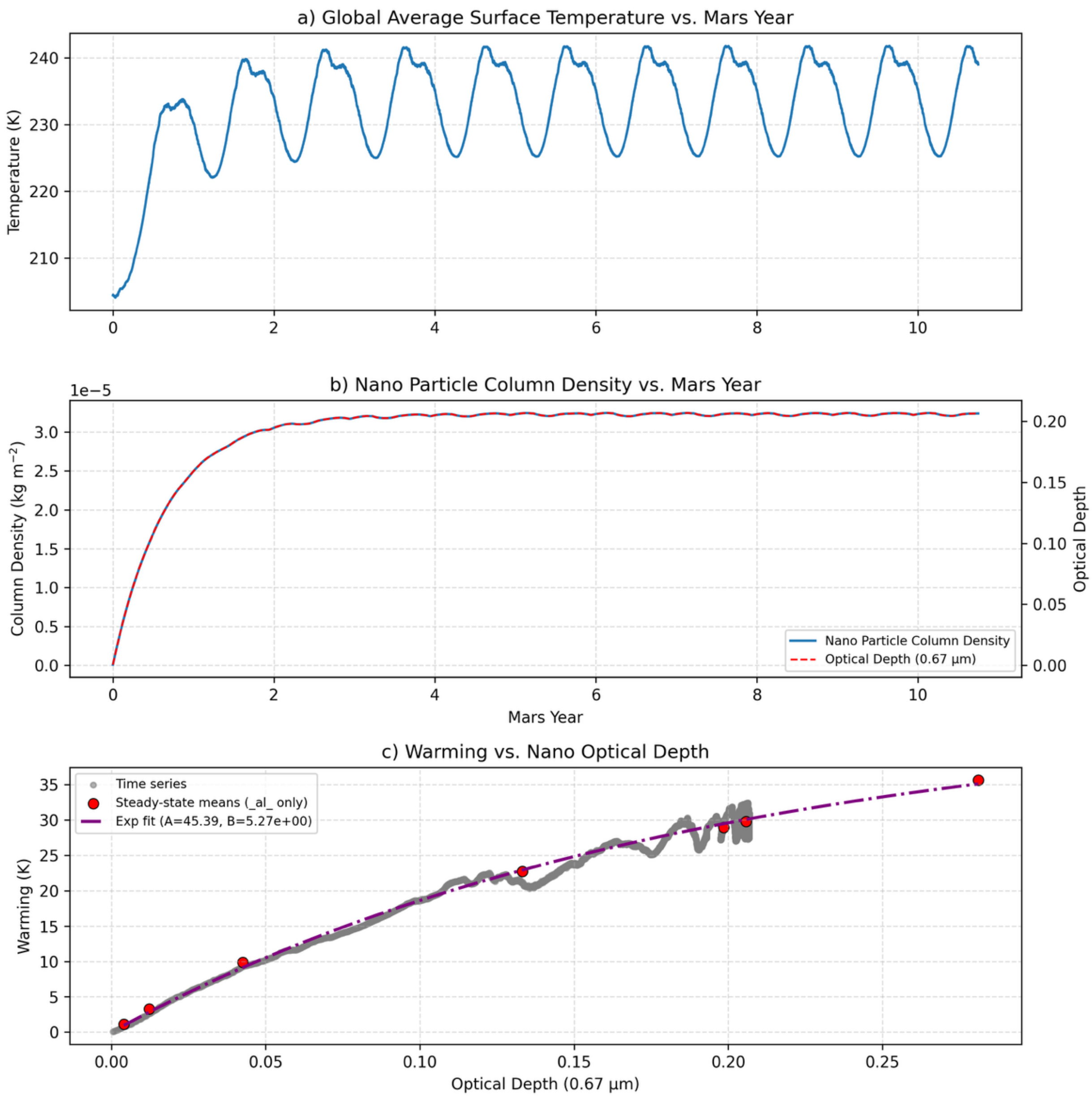

**Fig. S10. Transient and Steady State Temperature Response.** (a) Transient temperature response for 45 L/s Al rod loading (*run Cc17*). The global average for every sol is plotted. (b) Time history of global average nano particle column density and visible (0.67 µm) optical depth. (c) Relationship between surface temperature warming and visible optical depth ($\tau_{vis}$) during transient warming. Gray line shows the continuous trajectory during the simulation from (a) and (b). Warming is calculated by taking the instantaneous global average $T_{surf}$ and subtracting from it the global average $T_{surf}$ for that sol-of-year from the unperturbed case. Red dots show the steady state annual average warming and nano rod visible opacity from the Al cases listed in Table S3. The fit to the steady state values is the warming, $\Delta T = A(1-e^{-B\tau})$. Note

the fit is likely only valid to moderate optical depths where significant sunlight still penetrates to the surface.

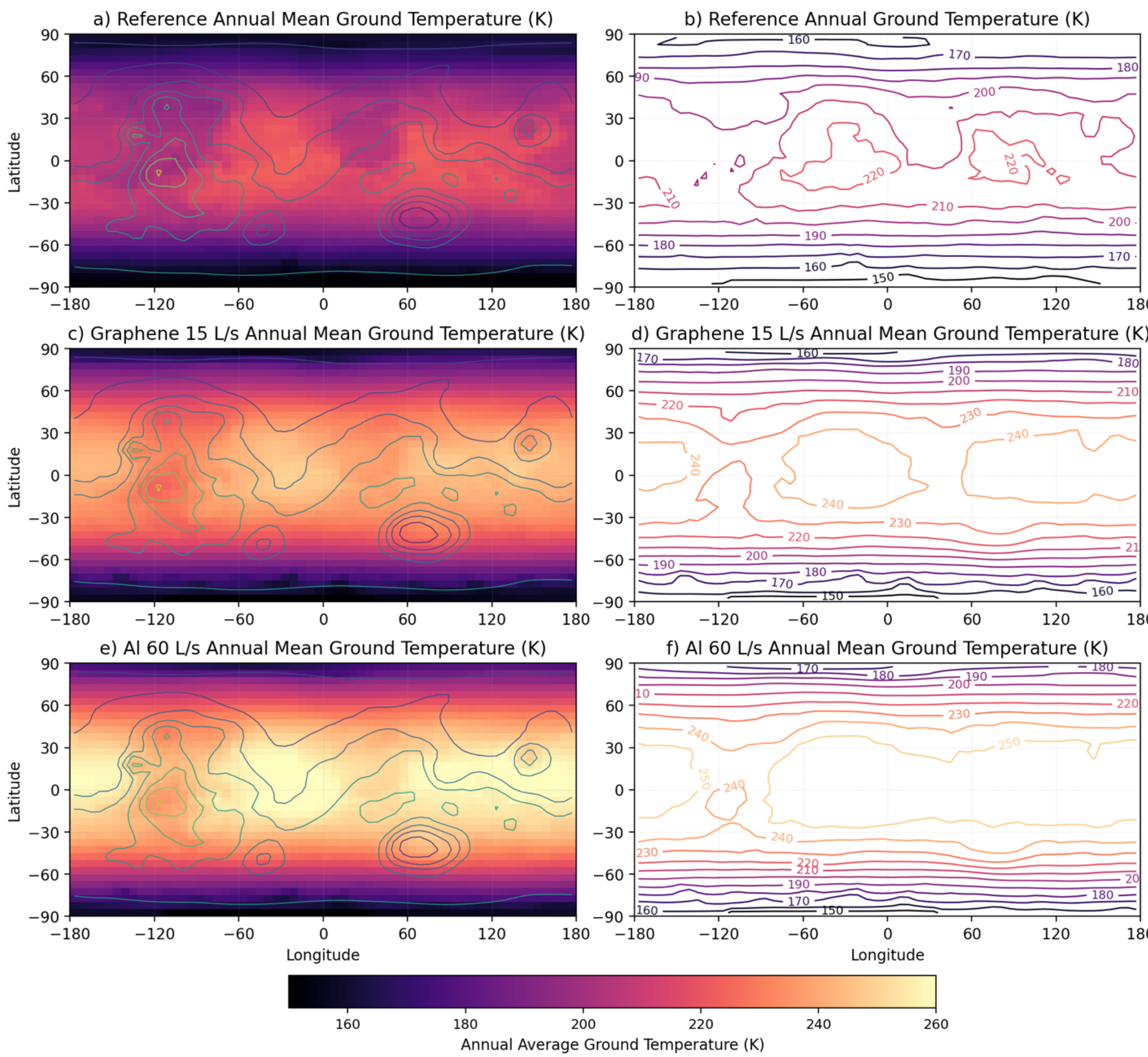

**Fig. S11. Global average temperatures (K) in steady state. (a,b) Control case (no warming). (c,d)** Graphene 15 L/s (*run Cc41*). **(e,f)** Al rods 60 L/s (*run Cc16*).

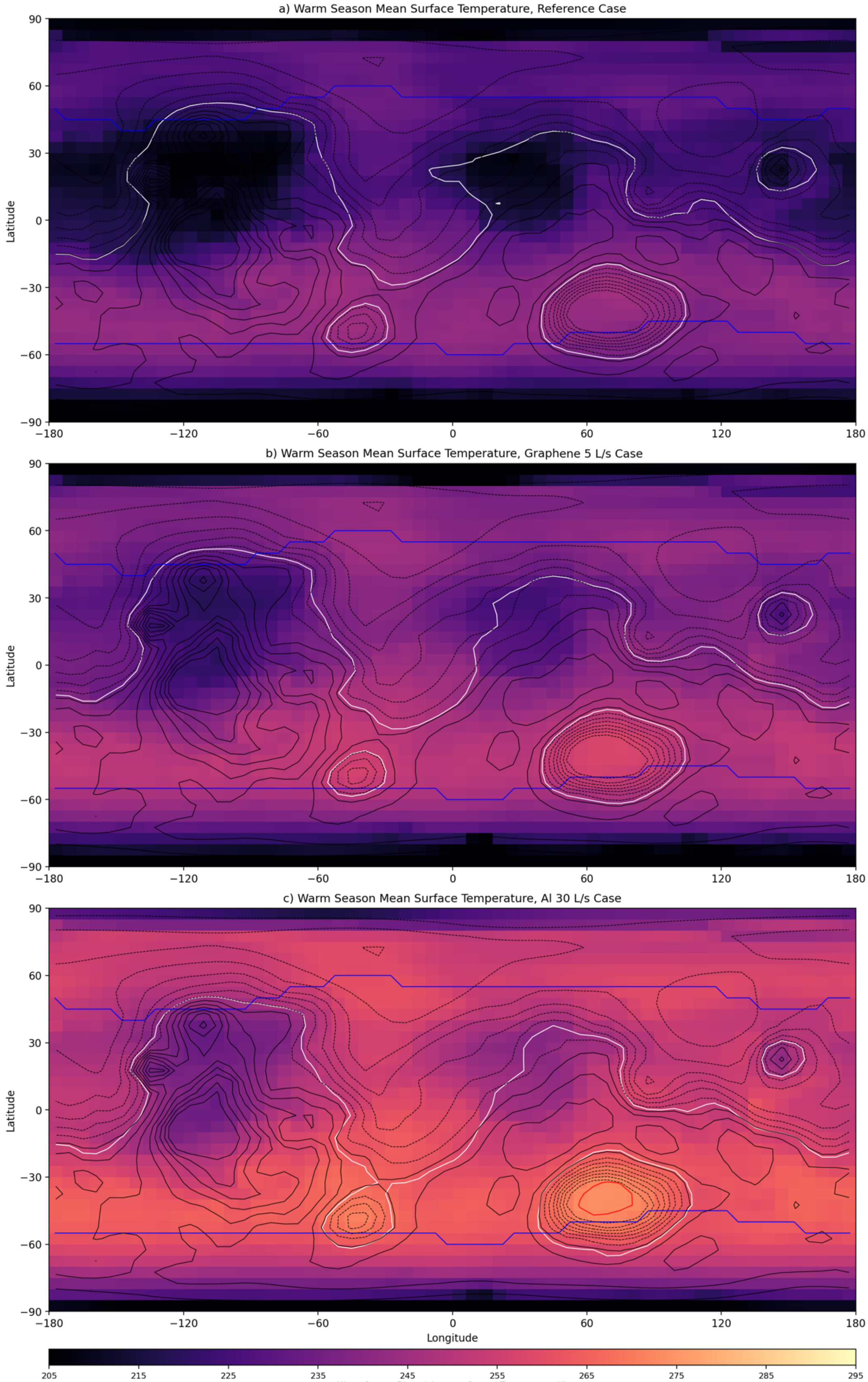

a) Warm Season Mean Surface Temperature, Reference Case
b) Warm Season Mean Surface Temperature, Graphene 5 L/s Case
c) Warm Season Mean Surface Temperature, Al 30 L/s Case
Latitude
Longitude
90
60
30
0
−30
−60
−90
−180
−120
−60
0
60
120
180
205
215
225
235
245
255
265
275
285
295
Warm Season Diurnal Average Ground Temperature (K)

**Fig. S12.**

**Warm-season average temperatures for control case, and sensitivity test for lower flux. (a)** Control case (no warming) (*run Cc8*). **(b)** Graphene disks, 5 L/s (*run Cc34*). **(c)** Al particles, 30 L/s (*run Cc29*). White contour corresponds to 610 Pa (~6 mbar) mean pressure level. Blue lines: Approximate latitudinal (equatorward) extent of ice at <1 m depths. Topographic contours correspond to elevations of –5 and –2 km (dashed), and 0, +2, and +5 km (solid).

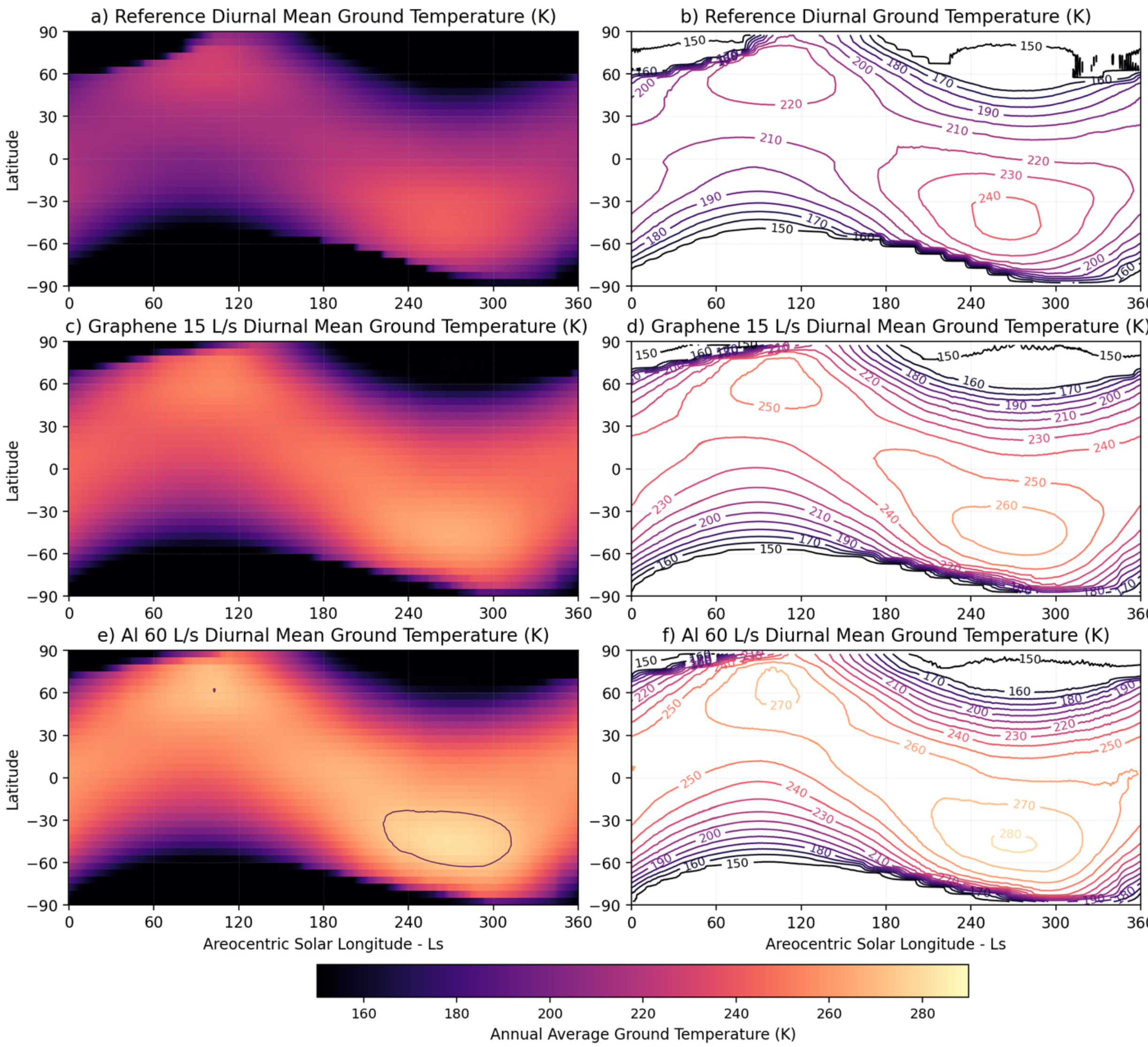

**Fig. S13. Seasonal variation of diurnally averaged surface temperature (K)**. **(a,b)** Control case (*run* Cc8). **(c,d)** Graphene case, 15 L/s (*run Cc41*). **(e,f)** Al particles, 60 L/s. (*run Cc16*). On panel (c) the contour line encloses the latitude-time region of diurnally averaged temperature about 273 K.

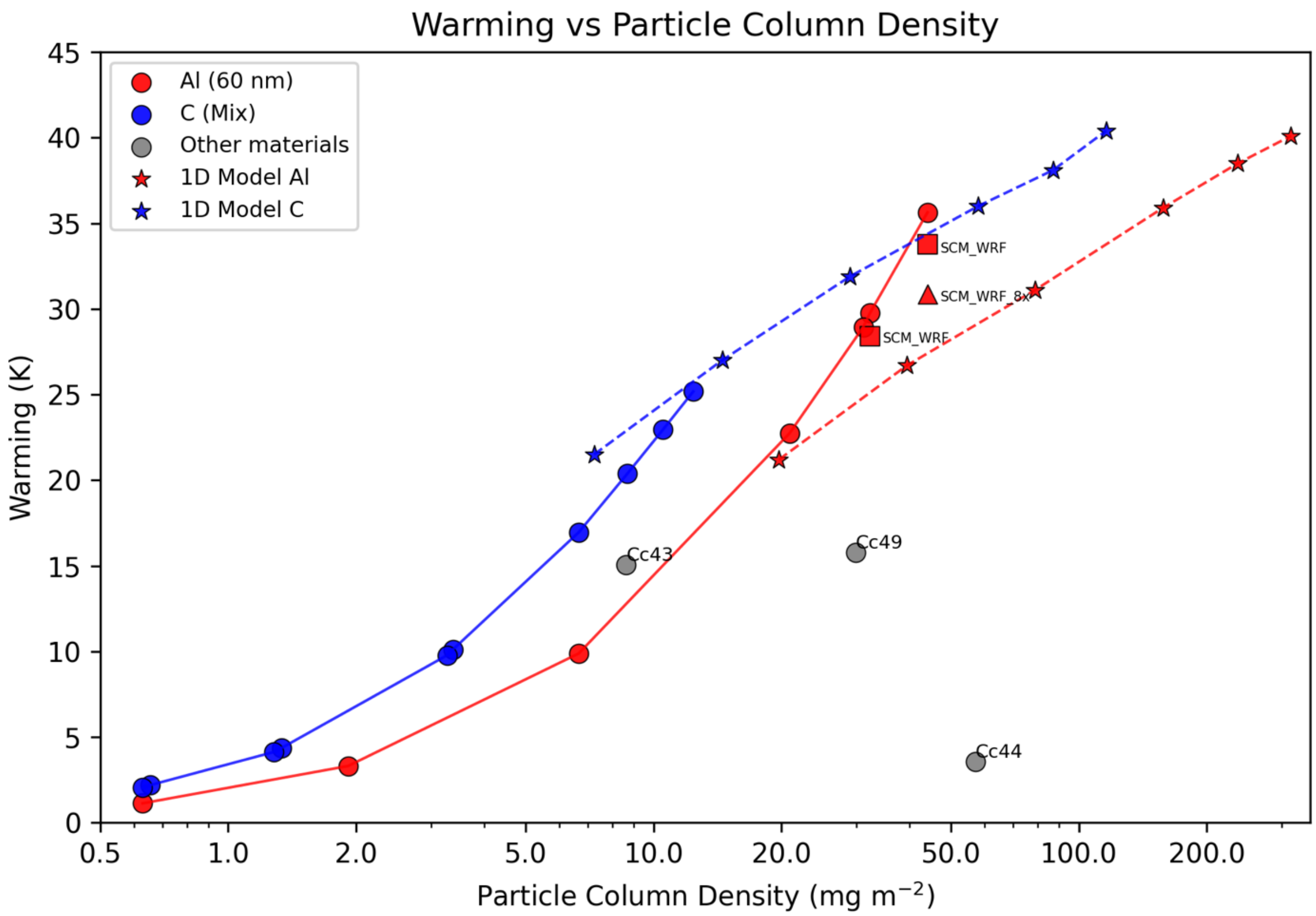

**Fig. S14. Global warming as a function of particle column density.** Comparison of 1D model output and WRF GCM 3D plume-tracking climate model output. Also shown are WRF Single Column Model (SCM) cases at GCM spectral resolution and at 8x spectral resolution (56 IR and 56 solar bands). Note that the 1D model no-warming temperature (the "zero" on this plot for the 1D runs) is 218 K, whereas the WRF 3D model no-warming temperature (the "zero" on this plot for the 3D runs) is 204.4 K. WRF SCM no-warming is 212.9 K (at both spectral resolutions).

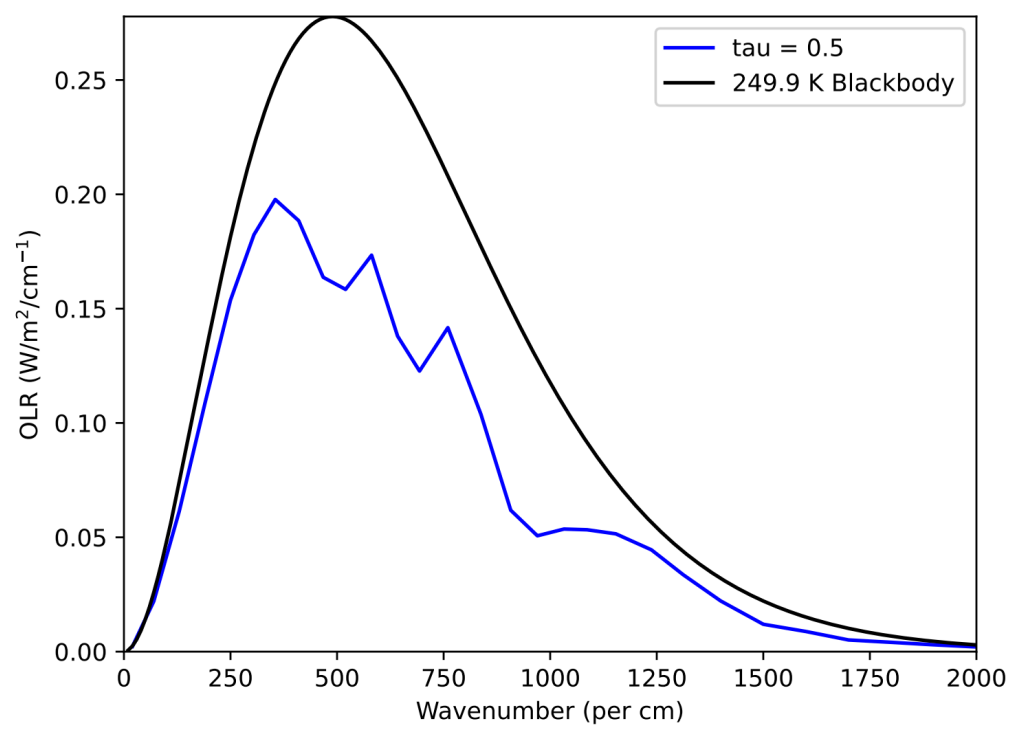

(a)

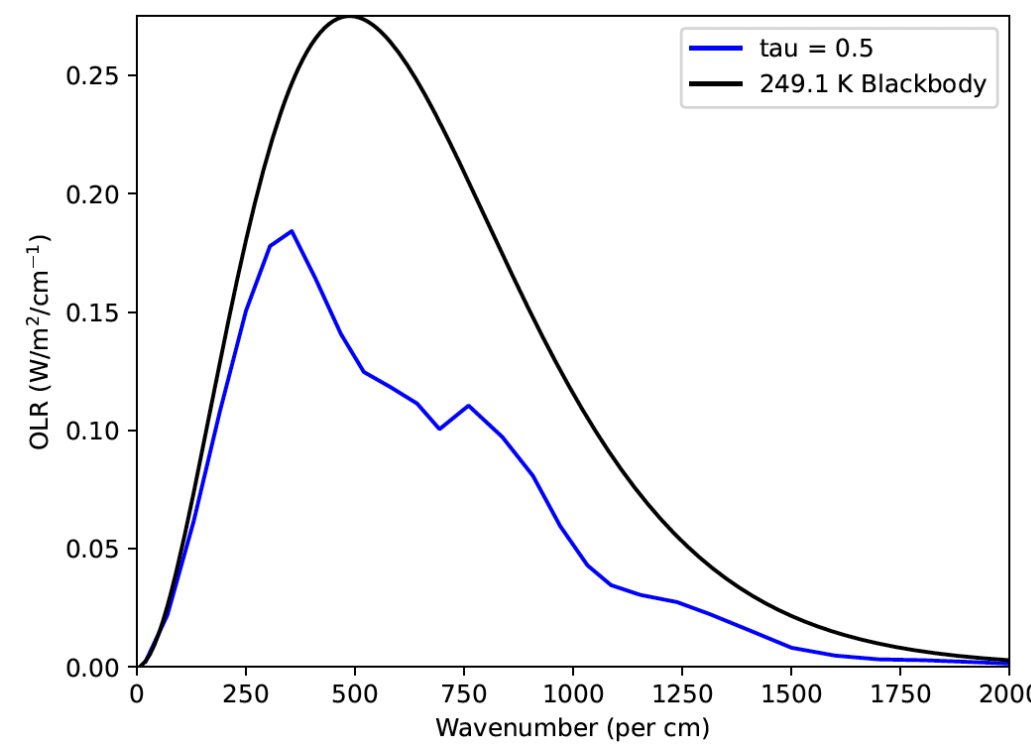

(b)

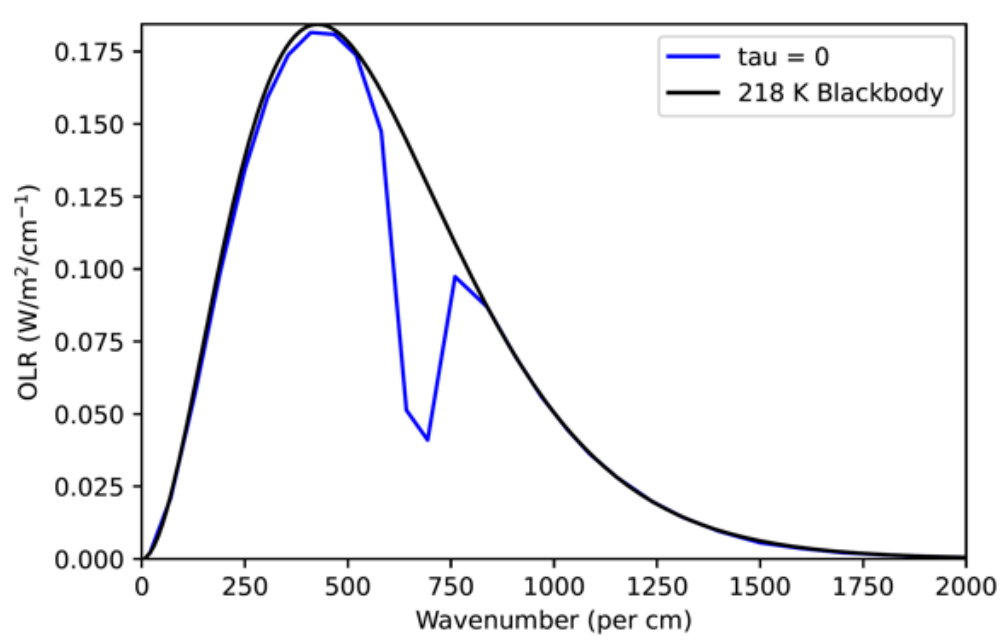

(c)

**Fig. S15. Examples of spectrally resolved 1D climate model output**. OLR = Outgoing Longwave Radiation. The blue curves correspond to the top-of-atmosphere thermal emission. The black curves correspond to the corresponding surface thermal emission. (a) Graphene ($\tau_{vis}$ = 0.5). (b) Al nanorods ($\tau_{vis}$ = 0.5). (c) No-warming case (this panel is reproduced from Ansari et al. 2024). 1D climate model output is summarized in Table S4.

## Supporting Tables.

| | Al rod (8 μm long × 60 nm × 60 nm) | | | | | | | |
|---|---|---|---|---|---|---|---|---|
| wavelength (μm) | scattering cross-section ($m^2$) | absorption cross-section ($m^2$) | asymmetry parameter, *g* | | wavelength (μm) | scattering cross-section ($m^2$) | absorption cross-section ($m^2$) | asymmetry parameter, *g* |
| 0.2 | 4.86E-13 | 4.46E-13 | 0.76182 | | 27.784 | 4.52E-13 | 4.82E-13 | 0.017968 |
| 0.56778 | 3.30E-13 | 2.00E-13 | 0.62875 | | 28.152 | 4.09E-13 | 4.80E-13 | 0.016856 |
| 0.93557 | 2.78E-13 | 1.24E-13 | 0.61122 | | 28.52 | 3.72E-13 | 4.76E-13 | 0.016044 |
| 1.3034 | 3.11E-13 | 1.49E-13 | 0.60043 | | 28.887 | 3.40E-13 | 4.70E-13 | 0.015525 |
| 1.6711 | 3.96E-13 | 2.02E-13 | 0.52437 | | 29.255 | 3.12E-13 | 4.60E-13 | 0.015278 |
| 2.0389 | 3.79E-13 | 1.44E-13 | 0.59837 | | 29.623 | 2.87E-13 | 4.47E-13 | 0.01527 |
| 2.4067 | 4.15E-13 | 1.68E-13 | 0.60509 | | 29.991 | 2.64E-13 | 4.30E-13 | 0.015457 |
| 2.7745 | 5.92E-13 | 2.31E-13 | 0.50739 | | 30.358 | 2.43E-13 | 4.11E-13 | 0.015788 |
| 3.1423 | 9.04E-13 | 4.58E-13 | 0.31906 | | 30.726 | 2.25E-13 | 3.90E-13 | 0.016207 |
| 3.5101 | 1.11E-12 | 5.63E-13 | 0.36581 | | 31.094 | 2.07E-13 | 3.68E-13 | 0.016665 |
| 3.8778 | 8.92E-13 | 5.21E-13 | 0.24 | | 31.462 | 1.91E-13 | 3.45E-13 | 0.017119 |
| 4.2456 | 9.22E-13 | 4.79E-13 | 0.34226 | | 31.829 | 1.77E-13 | 3.23E-13 | 0.017541 |
| 4.6134 | 1.67E-12 | 1.32E-12 | 0.27497 | | 32.197 | 1.63E-13 | 3.01E-13 | 0.01792 |
| 4.9812 | 8.40E-13 | 4.08E-13 | 0.22821 | | 32.565 | 1.51E-13 | 2.80E-13 | 0.018251 |
| 5.349 | 9.23E-13 | 4.01E-13 | 0.33662 | | 32.933 | 1.39E-13 | 2.61E-13 | 0.018537 |
| 5.7168 | 1.37E-12 | 5.22E-13 | 0.30799 | | 33.301 | 1.29E-13 | 2.44E-13 | 0.01878 |
| 6.0846 | 2.14E-12 | 1.62E-12 | 0.28861 | | 33.668 | 1.19E-13 | 2.30E-13 | 0.018971 |
| 6.4524 | 1.62E-12 | 1.07E-12 | 0.16687 | | 34.036 | 1.11E-13 | 2.17E-13 | 0.019094 |
| 6.8201 | 9.73E-13 | 5.21E-13 | 0.10957 | | 34.404 | 1.03E-13 | 2.08E-13 | 0.019122 |
| 7.1879 | 9.21E-13 | 3.83E-13 | 0.18737 | | 34.772 | 9.54E-14 | 2.00E-13 | 0.019019 |
| 7.5557 | 1.06E-12 | 5.03E-13 | 0.23115 | | 35.14 | 8.90E-14 | 1.96E-13 | 0.018752 |
| 7.9235 | 1.19E-12 | 5.50E-13 | 0.21333 | | 35.507 | 8.32E-14 | 1.93E-13 | 0.018297 |
| 8.2913 | 1.68E-12 | 7.03E-13 | 0.2203 | | 35.875 | 7.80E-14 | 1.92E-13 | 0.017641 |
| 8.6591 | 2.40E-12 | 1.31E-12 | 0.23074 | | 36.243 | 7.35E-14 | 1.94E-13 | 0.01679 |
| 9.0268 | 2.97E-12 | 2.17E-12 | 0.21556 | | 36.611 | 6.94E-14 | 1.97E-13 | 0.015761 |
| 9.3946 | 3.01E-12 | 2.61E-12 | 0.17771 | | 36.978 | 6.59E-14 | 2.01E-13 | 0.014583 |
| 9.7624 | 2.54E-12 | 2.17E-12 | 0.12243 | | 37.346 | 6.28E-14 | 2.07E-13 | 0.01329 |
| 10.13 | 1.90E-12 | 1.33E-12 | 0.055812 | | 37.714 | 6.01E-14 | 2.13E-13 | 0.011914 |
| 10.498 | 1.38E-12 | 7.55E-13 | -0.007209 | | 38.082 | 5.78E-14 | 2.21E-13 | 0.010486 |
| 10.866 | 1.11E-12 | 5.96E-13 | -0.038166 | | 38.45 | 5.58E-14 | 2.29E-13 | 0.0090313 |
| 11.234 | 1.04E-12 | 6.24E-13 | -0.030053 | | 38.817 | 5.41E-14 | 2.37E-13 | 0.0075691 |
| 11.601 | 1.07E-12 | 6.43E-13 | -0.010819 | | 39.185 | 5.26E-14 | 2.45E-13 | 0.0061152 |
| 11.969 | 1.13E-12 | 6.11E-13 | 0.0059081 | | 39.553 | 5.14E-14 | 2.53E-13 | 0.0046804 |
| 12.337 | 1.18E-12 | 5.62E-13 | 0.019553 | | 39.921 | 5.03E-14 | 2.60E-13 | 0.0032723 |
| 12.705 | 1.25E-12 | 5.44E-13 | 0.029777 | | 40.289 | 4.94E-14 | 2.68E-13 | 0.0018958 |
| 13.072 | 1.36E-12 | 5.91E-13 | 0.037433 | | 40.656 | 4.86E-14 | 2.74E-13 | 0.00055384 |
| 13.44 | 1.51E-12 | 7.15E-13 | 0.045057 | | 41.024 | 4.79E-14 | 2.81E-13 | -0.0007521 |
| 13.808 | 1.67E-12 | 8.97E-13 | 0.053475 | | 41.392 | 4.73E-14 | 2.86E-13 | -0.0020214 |
| 14.176 | 1.82E-12 | 1.09E-12 | 0.060761 | | 41.76 | 4.68E-14 | 2.91E-13 | -0.0032544 |
| 14.544 | 1.98E-12 | 1.24E-12 | 0.065881 | | 42.127 | 4.63E-14 | 2.94E-13 | -0.0044514 |
| 14.911 | 2.19E-12 | 1.34E-12 | 0.06997 | | 42.495 | 4.58E-14 | 2.97E-13 | -0.0056132 |
| 15.279 | 2.49E-12 | 1.40E-12 | 0.074165 | | 42.863 | 4.54E-14 | 2.99E-13 | -0.0067406 |
| 15.647 | 2.92E-12 | 1.50E-12 | 0.078465 | | 43.231 | 4.50E-14 | 3.00E-13 | -0.0078345 |
| 16.015 | 3.49E-12 | 1.71E-12 | 0.08234 | | 43.599 | 4.45E-14 | 3.00E-13 | -0.0088956 |
| 16.383 | 4.18E-12 | 2.09E-12 | 0.085375 | | 43.966 | 4.41E-14 | 3.00E-13 | -0.0099247 |
| 16.75 | 4.95E-12 | 2.68E-12 | 0.087416 | | 44.334 | 4.36E-14 | 2.98E-13 | -0.010922 |
| 17.118 | 5.71E-12 | 3.45E-12 | 0.088481 | | 44.702 | 4.32E-14 | 2.95E-13 | -0.011888 |
| 17.486 | 6.42E-12 | 4.34E-12 | 0.088671 | | 45.07 | 4.27E-14 | 2.92E-13 | -0.012824 |
| 17.854 | 7.00E-12 | 5.27E-12 | 0.088113 | | 45.438 | 4.22E-14 | 2.88E-13 | -0.013729 |
| 18.222 | 7.42E-12 | 6.14E-12 | 0.086934 | | 45.805 | 4.16E-14 | 2.83E-13 | -0.014603 |
| 18.589 | 7.66E-12 | 6.87E-12 | 0.085249 | | 46.173 | 4.10E-14 | 2.77E-13 | -0.015447 |
| 18.957 | 7.71E-12 | 7.39E-12 | 0.083162 | | 46.541 | 4.04E-14 | 2.70E-13 | -0.01626 |
| 19.325 | 7.58E-12 | 7.67E-12 | 0.080757 | | 46.909 | 3.98E-14 | 2.63E-13 | -0.017043 |
| 19.693 | 7.30E-12 | 7.70E-12 | 0.078108 | | 47.276 | 3.91E-14 | 2.56E-13 | -0.017795 |
| 20.06 | 6.90E-12 | 7.51E-12 | 0.075275 | | 47.644 | 3.84E-14 | 2.47E-13 | -0.018517 |
| 20.428 | 6.42E-12 | 7.11E-12 | 0.072306 | | 48.012 | 3.77E-14 | 2.39E-13 | -0.019208 |
| 20.796 | 5.87E-12 | 6.57E-12 | 0.069243 | | 48.38 | 3.69E-14 | 2.29E-13 | -0.019869 |
| 21.164 | 5.31E-12 | 5.92E-12 | 0.066116 | | 48.748 | 3.62E-14 | 2.20E-13 | -0.0205 |
| 21.531 | 4.74E-12 | 5.22E-12 | 0.062954 | | 49.115 | 3.54E-14 | 2.10E-13 | -0.021102 |
| 21.899 | 4.18E-12 | 4.51E-12 | 0.059778 | | 49.483 | 3.45E-14 | 2.00E-13 | -0.021675 |
| 22.267 | 3.66E-12 | 3.82E-12 | 0.056603 | | 49.851 | 3.37E-14 | 1.90E-13 | -0.022221 |
| 22.635 | 3.18E-12 | 3.19E-12 | 0.053444 | | 50.219 | 3.28E-14 | 1.79E-13 | -0.022739 |
| 23.003 | 2.75E-12 | 2.61E-12 | 0.050314 | | 50.587 | 3.20E-14 | 1.68E-13 | -0.023233 |
| 23.37 | 2.36E-12 | 2.12E-12 | 0.047221 | | 50.954 | 3.11E-14 | 1.57E-13 | -0.023701 |
| 23.738 | 2.02E-12 | 1.70E-12 | 0.044174 | | 51.322 | 3.02E-14 | 1.46E-13 | -0.024147 |
| 24.106 | 1.73E-12 | 1.36E-12 | 0.041183 | | 51.69 | 2.93E-14 | 1.35E-13 | -0.024571 |
| 24.474 | 1.48E-12 | 1.10E-12 | 0.038259 | | 52.058 | 2.84E-14 | 1.24E-13 | -0.024975 |
| 24.842 | 1.27E-12 | 8.94E-13 | 0.035412 | | 52.425 | 2.75E-14 | 1.13E-13 | -0.025359 |
| 25.209 | 1.09E-12 | 7.45E-13 | 0.032657 | | 52.793 | 2.65E-14 | 1.03E-13 | -0.025726 |
| 25.577 | 9.46E-13 | 6.40E-13 | 0.030013 | | 53.161 | 2.56E-14 | 9.17E-14 | -0.026075 |
| 25.945 | 8.23E-13 | 5.71E-13 | 0.027503 | | 53.529 | 2.47E-14 | 8.10E-14 | -0.026407 |
| 26.313 | 7.20E-13 | 5.27E-13 | 0.025154 | | 53.897 | 2.38E-14 | 7.04E-14 | -0.026724 |
| 26.68 | 6.34E-13 | 5.02E-13 | 0.022995 | | 54.264 | 2.29E-14 | 5.99E-14 | -0.027025 |
| 27.048 | 5.63E-13 | 4.89E-13 | 0.021058 | | 54.632 | 2.20E-14 | 4.97E-14 | -0.027311 |
| 27.416 | 5.03E-13 | 4.84E-13 | 0.019374 | | 55 | 2.11E-14 | 3.96E-14 | -0.027582 |

**Table S1.** Optical properties of engineered metal aerosol. Notes: Negative extinction cross sections are set to zero before use in the climate model. In code, albedos <0.01 and >0.99 are removed (i.e., albedo is bounded between those limits).

| | Graphene (1000 nm-diameter graphene disk) | | Graphene (250 nm-diameter graphene disk) | | | Graphene (1000 nm-diameter graphene disk) | | Graphene (250 nm-diameter graphene disk) | |
|---|---|---|---|---|---|---|---|---|---|
| λ (μm) | scattering cross-section ($m^2$) | absorption cross-section ($m^2$) | scattering cross-section ($m^2$) | absorption cross-section ($m^2$) | λ (μm) | scattering cross-section ($m^2$) | absorption cross-section ($m^2$) | scattering cross-section ($m^2$) | absorption cross-section ($m^2$) |
| 2.00E-07 | 4.51E-16 | 1.41E-14 | 2.47E-17 | 8.96E-16 | 7.14E-06 | 1.60E-17 | 9.59E-15 | 1.85E-20 | 5.16E-16 |
| 2.11E-07 | 5.47E-16 | 1.44E-14 | 3.04E-17 | 9.15E-16 | 7.22E-06 | 1.56E-17 | 9.35E-15 | 1.49E-20 | 5.19E-16 |
| 2.22E-07 | 7.93E-16 | 1.52E-14 | 4.47E-17 | 9.67E-16 | 7.30E-06 | 1.46E-17 | 9.14E-15 | 1.13E-20 | 5.22E-16 |
| 2.33E-07 | 1.48E-15 | 1.96E-14 | 8.41E-17 | 1.25E-15 | 7.38E-06 | 1.30E-17 | 8.98E-15 | 7.90E-21 | 5.26E-16 |
| 2.44E-07 | 2.86E-15 | 3.20E-14 | 1.64E-16 | 2.06E-15 | 7.46E-06 | 1.10E-17 | 8.88E-15 | 4.54E-21 | 5.29E-16 |
| 2.56E-07 | 4.41E-15 | 4.90E-14 | 2.53E-16 | 3.15E-15 | 7.53E-06 | 8.71E-18 | 8.84E-15 | 1.27E-21 | 5.33E-16 |
| 2.67E-07 | 5.08E-15 | 5.84E-14 | 2.90E-16 | 3.73E-15 | 7.61E-06 | 6.32E-18 | 8.85E-15 | -1.91E-21 | 5.37E-16 |
| 2.78E-07 | 4.57E-15 | 5.57E-14 | 2.61E-16 | 3.53E-15 | 7.69E-06 | 3.96E-18 | 8.93E-15 | -4.98E-21 | 5.41E-16 |
| 2.89E-07 | 3.44E-15 | 4.58E-14 | 1.97E-16 | 2.89E-15 | 7.77E-06 | 1.76E-18 | 9.05E-15 | -7.95E-21 | 5.45E-16 |
| 3.00E-07 | 2.36E-15 | 3.59E-14 | 1.35E-16 | 2.26E-15 | 7.84E-06 | -1.45E-19 | 9.21E-15 | -1.08E-20 | 5.49E-16 |
| 3.78E-07 | 3.05E-16 | 1.40E-14 | 1.54E-17 | 8.71E-16 | 7.92E-06 | -1.68E-18 | 9.39E-15 | -1.36E-20 | 5.54E-16 |
| 4.56E-07 | 2.37E-16 | 1.26E-14 | 1.01E-17 | 7.93E-16 | 8.00E-06 | -2.78E-18 | 9.59E-15 | -1.62E-20 | 5.58E-16 |
| 5.33E-07 | 1.94E-16 | 1.14E-14 | 6.75E-18 | 7.13E-16 | 8.50E-06 | 1.33E-16 | 4.28E-15 | 2.04E-18 | 1.18E-15 |
| 6.11E-07 | 1.60E-16 | 1.07E-14 | 4.70E-18 | 6.57E-16 | 9.06E-06 | 1.77E-16 | 1.33E-14 | 4.37E-18 | 3.39E-15 |
| 6.89E-07 | 1.37E-16 | 1.02E-14 | 3.48E-18 | 6.35E-16 | 9.61E-06 | 1.86E-16 | 1.98E-14 | 6.91E-18 | 4.74E-15 |
| 7.67E-07 | 1.24E-16 | 1.00E-14 | 2.70E-18 | 6.23E-16 | 1.02E-05 | 1.40E-16 | 2.40E-14 | 2.09E-17 | 1.72E-14 |
| 8.44E-07 | 1.19E-16 | 9.83E-15 | 2.16E-18 | 6.15E-16 | 1.07E-05 | 1.23E-16 | 8.39E-15 | 2.77E-17 | 2.71E-14 |
| 9.22E-07 | 1.17E-16 | 9.72E-15 | 1.80E-18 | 6.12E-16 | 1.13E-05 | 1.95E-16 | 2.00E-14 | 1.67E-17 | 1.64E-14 |
| 1.00E-06 | 1.16E-16 | 9.81E-15 | 1.52E-18 | 6.05E-16 | 1.18E-05 | 1.60E-16 | 1.42E-14 | 5.77E-18 | 5.40E-15 |
| 1.08E-06 | 1.14E-16 | 9.71E-15 | 1.31E-18 | 6.12E-16 | 1.24E-05 | 1.99E-16 | 1.18E-14 | 2.39E-18 | 3.13E-15 |
| 1.16E-06 | 1.12E-16 | 9.77E-15 | 1.12E-18 | 5.94E-16 | 1.29E-05 | 2.86E-16 | 2.78E-14 | 1.84E-18 | 2.73E-15 |
| 1.23E-06 | 1.11E-16 | 9.69E-15 | 1.01E-18 | 6.06E-16 | 1.35E-05 | 2.73E-16 | 3.09E-14 | 1.16E-18 | 1.37E-15 |
| 1.31E-06 | 1.06E-16 | 9.64E-15 | 8.76E-19 | 6.07E-16 | 1.40E-05 | 2.55E-16 | 2.28E-14 | 5.64E-19 | 5.64E-16 |
| 1.39E-06 | 1.01E-16 | 9.42E-15 | 7.69E-19 | 5.78E-16 | 1.46E-05 | 3.49E-16 | 3.03E-14 | 4.07E-19 | 7.32E-16 |
| 1.47E-06 | 9.86E-17 | 9.59E-15 | 7.06E-19 | 5.93E-16 | 1.51E-05 | 4.89E-16 | 5.42E-14 | 4.30E-19 | 9.69E-16 |
| 1.54E-06 | 9.47E-17 | 9.68E-15 | 6.49E-19 | 6.18E-16 | 1.57E-05 | 5.58E-16 | 7.31E-14 | 3.72E-19 | 7.68E-16 |
| 1.62E-06 | 8.96E-17 | 9.58E-15 | 5.78E-19 | 6.01E-16 | 1.63E-05 | 5.60E-16 | 7.80E-14 | 2.32E-19 | 3.53E-16 |
| 1.70E-06 | 8.38E-17 | 9.35E-15 | 5.09E-19 | 5.66E-16 | 1.68E-05 | 6.10E-16 | 8.30E-14 | 1.16E-19 | 9.95E-17 |
| 1.78E-06 | 8.07E-17 | 9.40E-15 | 4.59E-19 | 5.62E-16 | 1.74E-05 | 8.17E-16 | 1.10E-13 | 8.08E-20 | 1.23E-16 |
| 1.86E-06 | 8.02E-17 | 9.98E-15 | 4.31E-19 | 5.90E-16 | 1.79E-05 | 1.19E-15 | 1.70E-13 | 1.03E-19 | 2.94E-16 |
| 1.93E-06 | 7.28E-17 | 9.67E-15 | 4.13E-19 | 6.22E-16 | 1.85E-05 | 1.65E-15 | 2.55E-13 | 1.33E-19 | 4.32E-16 |
| 2.01E-06 | 6.98E-17 | 9.38E-15 | 3.92E-19 | 6.30E-16 | 1.90E-05 | 2.07E-15 | 3.45E-13 | 1.36E-19 | 4.43E-16 |
| 2.09E-06 | 6.83E-17 | 1.01E-14 | 3.63E-19 | 6.12E-16 | 1.96E-05 | 2.34E-15 | 4.17E-13 | 1.09E-19 | 3.36E-16 |
| 2.17E-06 | 6.18E-17 | 9.22E-15 | 3.29E-19 | 5.81E-16 | 2.01E-05 | 2.42E-15 | 4.58E-13 | 6.63E-20 | 1.80E-16 |
| 2.24E-06 | 6.19E-17 | 1.01E-14 | 2.93E-19 | 5.55E-16 | 2.07E-05 | 2.31E-15 | 4.63E-13 | 2.74E-20 | 4.76E-17 |
| 2.32E-06 | 5.70E-17 | 9.28E-15 | 2.63E-19 | 5.46E-16 | 2.12E-05 | 2.05E-15 | 4.36E-13 | 4.14E-21 | 4.34E-17 |
| 2.40E-06 | 5.35E-17 | 9.93E-15 | 2.40E-19 | 5.56E-16 | 2.18E-05 | 1.72E-15 | 3.85E-13 | -5.57E-22 | 4.34E-17 |
| 2.48E-06 | 5.51E-17 | 9.78E-15 | 2.25E-19 | 5.79E-16 | 2.23E-05 | 1.36E-15 | 3.23E-13 | 9.05E-21 | 3.61E-17 |
| 2.56E-06 | 4.63E-17 | 9.26E-15 | 2.16E-19 | 6.05E-16 | 2.29E-05 | 1.03E-15 | 2.59E-13 | 2.55E-20 | 1.12E-16 |
| 2.63E-06 | 4.85E-17 | 1.02E-14 | 2.12E-19 | 6.27E-16 | 2.34E-05 | 7.54E-16 | 2.00E-13 | 4.16E-20 | 1.83E-16 |
| 2.71E-06 | 4.80E-17 | 9.56E-15 | 2.10E-19 | 6.40E-16 | 2.40E-05 | 5.38E-16 | 1.50E-13 | 5.22E-20 | 2.30E-16 |
| 2.79E-06 | 3.97E-17 | 9.22E-15 | 2.08E-19 | 6.42E-16 | 2.46E-05 | 3.85E-16 | 1.12E-13 | 5.51E-20 | 2.44E-16 |
| 2.87E-06 | 4.07E-17 | 1.02E-14 | 2.04E-19 | 6.33E-16 | 2.51E-05 | 2.85E-16 | 8.45E-14 | 5.06E-20 | 2.29E-16 |
| 2.94E-06 | 4.37E-17 | 9.97E-15 | 1.98E-19 | 6.17E-16 | 2.57E-05 | 2.24E-16 | 6.65E-14 | 4.03E-20 | 1.90E-16 |
| 3.02E-06 | 3.76E-17 | 9.11E-15 | 1.90E-19 | 5.96E-16 | 2.62E-05 | 1.92E-16 | 5.56E-14 | 2.68E-20 | 1.38E-16 |
| 3.10E-06 | 3.20E-17 | 9.51E-15 | 1.79E-19 | 5.74E-16 | 2.68E-05 | 1.76E-16 | 4.98E-14 | 1.27E-20 | 8.20E-17 |
| 3.18E-06 | 3.46E-17 | 1.03E-14 | 1.66E-19 | 5.54E-16 | 2.73E-05 | 1.67E-16 | 4.71E-14 | 1.00E-22 | 3.15E-17 |
| 3.26E-06 | 3.77E-17 | 1.00E-14 | 1.52E-19 | 5.39E-16 | 2.79E-05 | 1.61E-16 | 4.60E-14 | -9.61E-21 | 4.34E-17 |
| 3.33E-06 | 3.37E-17 | 9.21E-15 | 1.38E-19 | 5.30E-16 | 2.84E-05 | 1.52E-16 | 4.53E-14 | -1.57E-20 | 4.34E-17 |
| 3.41E-06 | 2.71E-17 | 9.17E-15 | 1.24E-19 | 5.26E-16 | 2.90E-05 | 1.41E-16 | 4.44E-14 | -1.81E-20 | 4.34E-17 |
| 3.49E-06 | 2.54E-17 | 9.92E-15 | 1.11E-19 | 5.28E-16 | 2.95E-05 | 1.26E-16 | 4.28E-14 | -1.72E-20 | 4.34E-17 |
| 3.57E-06 | 2.91E-17 | 1.04E-14 | 9.92E-20 | 5.36E-16 | 3.01E-05 | 1.08E-16 | 4.05E-14 | -1.36E-20 | 4.34E-17 |
| 3.64E-06 | 3.20E-17 | 1.01E-14 | 8.90E-20 | 5.47E-16 | 3.06E-05 | 8.86E-17 | 3.75E-14 | -8.20E-21 | 4.34E-17 |
| 3.72E-06 | 2.97E-17 | 9.34E-15 | 8.04E-20 | 5.61E-16 | 3.12E-05 | 6.90E-17 | 3.40E-14 | -1.77E-21 | 1.17E-17 |
| 3.80E-06 | 2.40E-17 | 9.02E-15 | 7.36E-20 | 5.77E-16 | 3.18E-05 | 5.02E-17 | 3.02E-14 | 4.91E-21 | 3.81E-17 |
| 3.88E-06 | 1.95E-17 | 9.39E-15 | 6.85E-20 | 5.94E-16 | 3.23E-05 | 3.33E-17 | 2.63E-14 | 1.12E-20 | 6.38E-17 |
| 3.96E-06 | 1.94E-17 | 1.01E-14 | 6.49E-20 | 6.10E-16 | 3.29E-05 | 1.89E-17 | 2.24E-14 | 1.67E-20 | 8.69E-17 |
| 4.03E-06 | 2.27E-17 | 1.05E-14 | 6.28E-20 | 6.25E-16 | 3.34E-05 | 7.42E-18 | 1.88E-14 | 2.10E-20 | 1.06E-16 |
| 4.11E-06 | 2.61E-17 | 1.03E-14 | 6.19E-20 | 6.39E-16 | 3.40E-05 | -1.06E-18 | 1.56E-14 | 2.39E-20 | 1.20E-16 |
| 4.19E-06 | 2.66E-17 | 9.70E-15 | 6.21E-20 | 6.50E-16 | 3.45E-05 | -6.59E-18 | 1.28E-14 | 2.55E-20 | 1.30E-16 |
| 4.27E-06 | 2.36E-17 | 9.16E-15 | 6.32E-20 | 6.59E-16 | 3.51E-05 | -9.40E-18 | 1.05E-14 | 2.58E-20 | 1.34E-16 |
| 4.34E-06 | 1.87E-17 | 8.98E-15 | 6.49E-20 | 6.66E-16 | 3.56E-05 | -9.84E-18 | 8.62E-15 | 2.48E-20 | 1.33E-16 |
| 4.42E-06 | 1.44E-17 | 9.25E-15 | 6.70E-20 | 6.70E-16 | 3.62E-05 | -8.31E-18 | 7.25E-15 | 2.29E-20 | 1.28E-16 |
| 4.50E-06 | 1.28E-17 | 9.78E-15 | 6.95E-20 | 6.72E-16 | 3.67E-05 | -5.25E-18 | 6.30E-15 | 2.01E-20 | 1.20E-16 |
| 4.58E-06 | 1.41E-17 | 1.03E-14 | 7.21E-20 | 6.71E-16 | 3.73E-05 | -1.09E-18 | 5.74E-15 | 1.67E-20 | 1.08E-16 |
| 4.66E-06 | 1.73E-17 | 1.05E-14 | 7.47E-20 | 6.68E-16 | 3.78E-05 | 3.75E-18 | 5.51E-15 | 1.29E-20 | 9.51E-17 |
| 4.73E-06 | 2.05E-17 | 1.04E-14 | 7.72E-20 | 6.64E-16 | 3.84E-05 | 8.92E-18 | 5.54E-15 | 8.97E-21 | 8.05E-17 |
| 4.81E-06 | 2.23E-17 | 9.98E-15 | 7.96E-20 | 6.58E-16 | 3.89E-05 | 1.41E-17 | 5.77E-15 | 4.97E-21 | 6.53E-17 |
| 4.89E-06 | 2.17E-17 | 9.49E-15 | 8.17E-20 | 6.51E-16 | 3.95E-05 | 1.90E-17 | 6.16E-15 | 1.09E-21 | 5.00E-17 |
| 4.97E-06 | 1.91E-17 | 9.10E-15 | 8.35E-20 | 6.42E-16 | 4.01E-05 | 2.35E-17 | 6.65E-15 | -2.54E-21 | 3.52E-17 |
| 5.04E-06 | 1.52E-17 | 8.94E-15 | 8.49E-20 | 6.33E-16 | 4.06E-05 | 2.74E-17 | 7.20E-15 | -5.84E-21 | 2.12E-17 |
| 5.12E-06 | 1.13E-17 | 9.05E-15 | 8.60E-20 | 6.23E-16 | 4.12E-05 | 3.05E-17 | 7.76E-15 | -8.74E-21 | 8.43E-18 |
| 5.20E-06 | 8.33E-18 | 9.36E-15 | 8.66E-20 | 6.13E-16 | 4.17E-05 | 3.29E-17 | 8.31E-15 | -1.12E-20 | 4.34E-17 |
| 5.28E-06 | 7.08E-18 | 9.78E-15 | 8.68E-20 | 6.03E-16 | 4.23E-05 | 3.45E-17 | 8.82E-15 | -1.32E-20 | 4.34E-17 |
| 5.36E-06 | 7.65E-18 | 1.02E-14 | 8.67E-20 | 5.93E-16 | 4.28E-05 | 3.54E-17 | 9.28E-15 | -1.47E-20 | 4.34E-17 |
| 5.43E-06 | 9.68E-18 | 1.04E-14 | 8.61E-20 | 5.83E-16 | 4.34E-05 | 3.54E-17 | 9.66E-15 | -1.57E-20 | 4.34E-17 |
| 5.51E-06 | 1.25E-17 | 1.05E-14 | 8.51E-20 | 5.74E-16 | 4.39E-05 | 3.48E-17 | 9.97E-15 | -1.63E-20 | 4.34E-17 |
| 5.59E-06 | 1.54E-17 | 1.04E-14 | 8.38E-20 | 5.65E-16 | 4.45E-05 | 3.36E-17 | 1.02E-14 | -1.64E-20 | 4.34E-17 |
| 5.67E-06 | 1.75E-17 | 1.02E-14 | 8.22E-20 | 5.56E-16 | 4.50E-05 | 3.18E-17 | 1.03E-14 | -1.62E-20 | 4.34E-17 |
| 5.74E-06 | 1.86E-17 | 9.83E-15 | 8.02E-20 | 5.48E-16 | 4.56E-05 | 2.95E-17 | 1.04E-14 | -1.57E-20 | 4.34E-17 |
| 5.82E-06 | 1.82E-17 | 9.47E-15 | 7.79E-20 | 5.41E-16 | 4.61E-05 | 2.68E-17 | 1.03E-14 | -1.48E-20 | 4.34E-17 |
| 5.90E-06 | 1.66E-17 | 9.17E-15 | 7.54E-20 | 5.34E-16 | 4.67E-05 | 2.39E-17 | 1.02E-14 | -1.37E-20 | 4.34E-17 |
| 5.98E-06 | 1.40E-17 | 8.97E-15 | 7.26E-20 | 5.28E-16 | 4.73E-05 | 2.06E-17 | 1.00E-14 | -1.24E-20 | 4.34E-17 |
| 6.06E-06 | 1.09E-17 | 8.90E-15 | 6.96E-20 | 5.23E-16 | 4.78E-05 | 1.73E-17 | 9.76E-15 | -1.09E-20 | 4.34E-17 |
| 6.13E-06 | 7.76E-18 | 8.97E-15 | 6.64E-20 | 5.19E-16 | 4.84E-05 | 1.38E-17 | 9.45E-15 | -9.30E-21 | 4.34E-17 |
| 6.21E-06 | 4.98E-18 | 9.15E-15 | 6.31E-20 | 5.15E-16 | 4.89E-05 | 1.03E-17 | 9.10E-15 | -7.61E-21 | 4.34E-17 |
| 6.29E-06 | 2.92E-18 | 9.41E-15 | 5.96E-20 | 5.12E-16 | 4.95E-05 | 6.82E-18 | 8.70E-15 | -5.88E-21 | 4.34E-17 |
| 6.37E-06 | 1.80E-18 | 9.72E-15 | 5.61E-20 | 5.10E-16 | 5.00E-05 | 3.42E-18 | 8.27E-15 | -4.12E-21 | 4.34E-17 |
| 6.44E-06 | 1.69E-18 | 1.00E-14 | 5.24E-20 | 5.08E-16 | 5.06E-05 | 1.34E-19 | 7.82E-15 | -2.37E-21 | 4.34E-17 |
| 6.52E-06 | 2.52E-18 | 1.03E-14 | 4.86E-20 | 5.07E-16 | 5.11E-05 | -3.00E-18 | 7.34E-15 | -6.62E-22 | 2.26E-18 |
| 6.60E-06 | 4.12E-18 | 1.04E-14 | 4.49E-20 | 5.06E-16 | 5.17E-05 | -5.95E-18 | 6.86E-15 | 9.96E-22 | 8.07E-18 |
| 6.68E-06 | 6.26E-18 | 1.05E-14 | 4.11E-20 | 5.06E-16 | 5.22E-05 | -8.70E-18 | 6.37E-15 | 2.58E-21 | 1.38E-17 |
| 6.76E-06 | 8.65E-18 | 1.05E-14 | 3.72E-20 | 5.07E-16 | 5.28E-05 | -1.12E-17 | 5.88E-15 | 4.08E-21 | 1.94E-17 |
| 6.83E-06 | 1.10E-17 | 1.05E-14 | 3.34E-20 | 5.08E-16 | 5.33E-05 | -1.35E-17 | 5.39E-15 | 5.48E-21 | 2.48E-17 |
| 6.91E-06 | 1.31E-17 | 1.03E-14 | 2.96E-20 | 5.09E-16 | 5.39E-05 | -1.56E-17 | 4.91E-15 | 6.77E-21 | 2.99E-17 |
| 6.99E-06 | 1.47E-17 | 1.01E-14 | 2.59E-20 | 5.11E-16 | 5.44E-05 | -1.74E-17 | 4.44E-15 | 7.94E-21 | 3.48E-17 |
| 7.07E-06 | 1.57E-17 | 9.84E-15 | 2.21E-20 | 5.14E-16 | 5.50E-05 | -1.89E-17 | 3.99E-15 | 9.00E-21 | 3.93E-17 |

**Table S2.** Optical properties of modeled graphene disks. Asymmetry parameter is close to zero. (Negative cross-sections are set to zero before use in the climate model).

| Material | run_name | Release Site | Release Rate (L/s) | Warming (K) | Average Ground Temperature (K) | Particle Column Density (kg/m2) | Average Surface Pressure (Pa) | Warm Season Temperature in 47.5S Band (K) | Steady State Annual Mean Particle Visible Optical Depth |
|---|---|---|---|---|---|---|---|---|---|
| | reference - Cc8 | | | 0 | 204.4 | 0 | 625.67 | 241.8 | 0 |
| Al (60 nm) | Cc32 | Equatorial | 1 | 1.13 | 205.50 | 6.29E-07 | 626.85 | 242.87 | 4.01E-03 |
| Al (60 nm) | Cc31 | Equatorial | 3 | 3.31 | 207.68 | 1.91E-06 | 629.28 | 244.96 | 1.22E-02 |
| Al (60 nm) | Cc30 | Equatorial | 10 | 9.88 | 214.25 | 6.67E-06 | 639.64 | 251.61 | 4.25E-02 |
| Al (60 nm) | Cc29 | Equatorial | 30 | 22.74 | 227.11 | 2.09E-05 | 667.30 | 265.63 | 1.33E-01 |
| Al (60 nm) | Cc17 | Equatorial | 45 | 29.76 | 234.13 | 3.23E-05 | 680.13 | 273.49 | 2.06E-01 |
| Al (60 nm) | Cc16 | Equatorial | 60 | 35.64 | 240.00 | 4.41E-05 | 689.01 | 280.29 | 2.81E-01 |
| C (Mix) | Cc36 | Equatorial | 1 | 2.20 | 206.57 | 6.56E-07 | 625.26 | 243.86 | 9.47E-03 |
| C (Mix) | Cc35 | Equatorial | 2 | 4.37 | 208.73 | 1.33E-06 | 627.11 | 245.95 | 1.93E-02 |
| C (Mix) | Cc34 | Equatorial | 5 | 10.12 | 214.48 | 3.38E-06 | 636.93 | 251.55 | 4.88E-02 |
| C (Mix) | Cc33 | Equatorial | 12.5 | 20.38 | 224.75 | 8.68E-06 | 662.42 | 262.44 | 1.25E-01 |
| C (Mix) | Cc41 | Equatorial | 15 | 22.96 | 227.32 | 1.05E-05 | 666.81 | 265.34 | 1.52E-01 |
| C (Mix) | Cc42 | Equatorial | 17.5 | 25.18 | 229.54 | 1.24E-05 | 671.94 | 267.88 | 1.79E-01 |
| Al (60 nm) | Cc15 | Mid-Lat | 45 | 28.94 | 233.31 | 3.11E-05 | 679.89 | 272.41 | 1.99E-01 |
| C (Mix) | Cc9 | Mid-Lat | 1 | 2.04 | 206.40 | 6.28E-07 | 623.58 | 243.67 | 9.07E-03 |
| C (Mix) | Cc12 | Mid-Lat | 2 | 4.13 | 208.50 | 1.28E-06 | 624.74 | 245.58 | 1.85E-02 |
| C (Mix) | Cc10 | Mid-Lat | 5 | 9.78 | 214.15 | 3.28E-06 | 635.70 | 251.02 | 4.73E-02 |
| C (Mix) | Cc13 | Mid-Lat | 10 | 16.96 | 221.32 | 6.67E-06 | 656.54 | 258.62 | 9.64E-02 |
| C (1000 nm) | Cc43 | Equatorial | 15 | 15.05 | 219.42 | 8.63E-06 | 642.58 | 255.82 | 7.48E-02 |
| C (250 nm) | Cc44 | Equatorial | 100 | 3.58 | 207.94 | 5.71E-05 | 632.85 | 246.00 | 3.05E-02 |
| Al (160 nm) | Cc49 | Equatorial | 45 | 15.76 | 220.13 | 2.99E-05 | 656.30 | 258.52 | 1.43E-01 |
| | | | | | | | | | |

**Table S3.** Summary of 3D plume-tracking climate model output for runs to steady state, including sensitivity tests. For Release Site: "Mid-Lat" is Arcadia, 40°N, 202°E. "Equatorial" is Elysium, 0°N, 135°E. “C (mix)” refers to a mix of 16:1 (by number) 250 nm-diameter graphene disks to 1000 nm-graphene disks. For the last three rows, “Al (160 nm)” refers to the 9 μm × 160 nm × 160 nm particles modeled in Ansari et al. (2024), and “C (1000 nm)” and “C (250 nm)” refer to the sensitivity tests shown in Fig. S5. Warm-season temperature is defined as the average temperature during the warmest 70-sol period of the year.

| Optical depth @ 0.67 µm, $\tau_{vis}$ | Column mass (mg/m$^2$) (Graphene) | $T_{surf}$ (K) | Albedo | OLR (W/m$^2$) |
|---|---|---|---|---|
| 0 | 0 | 218.0 | 0.2178 | 114.10 |
| 0.125 | 7.25 | 239.5 | 0.1301 | 127.39 |
| 0.25 | 14.5 | 245.0 | 0.0804 | 134.58 |
| 0.5 | 29 | 249.9 | 0.0335 | 142.54 |
| 1 | 58 | 254.0 | 0.0087 | 145.52 |
| 1.5 | 86.9 | 256.1 | 0.0048 | 145.52 |
| 2 | 115.9 | 258.4 | 0.0041 | 147.86 |
| Optical depth @ 0.67 µm, $\tau_{vis}$ | Column mass (mg/m$^2$) (Al rods) | $T_{surf}$ (K) | PALB | OLR (W/m$^2$) |
| 0 | 0 | 218.0 | 0.2178 | 114.1 |
| 0.125 | 19.6 | 239.2 | 0.1993 | 117.62 |
| 0.25 | 39.3 | 244.7 | 0.1861 | 119.96 |
| 0.5 | 78.5 | 249.1 | 0.1693 | 121.66 |
| 1 | 157.1 | 253.9 | 0.1531 | 124.76 |
| 1.5 | 235.7 | 256.5 | 0.1468 | 125.37 |
| 2 | 314.2 | 258.1 | 0.1444 | 125.152 |

**Table S4.** Summary of 1-D model output. PALB = planetary albedo. OLR = Outgoing Longwave Radiation.